\documentclass[journal=jctcce,manuscript=article,layout=traditional]{achemso}
\setkeys{acs}{articletitle = true}

\usepackage[version=3]{mhchem} 
\usepackage[T1]{fontenc}       
\usepackage{xcolor}
\usepackage{amsmath}
\usepackage{amsfonts}
\usepackage{bm}
\usepackage{physics}
\usepackage{booktabs}
\usepackage{siunitx}
\usepackage[subrefformat=parens,labelformat=parens]{subcaption}
\usepackage{xr}
\SectionNumbersOff

\newcommand{\mcl}[1]{\mathcal{#1}}

\newcommand{\del}[1]{\delta_{#1}}

\newcommand{\ECC}{\mathrm{E_{CC}}}
\newcommand{\CC}{\mathrm{CC}}
\newcommand{\tCC}{\mathrm{\tilde{CC}}}
\newcommand{\R}{\phi_0}

\newcommand{\eps}{\varepsilon}
\newcommand{\mus}{\mu_1}
\newcommand{\mud}{\mu_2}
\newcommand{\mut}{\mu^a_3}
\newcommand{\nus}{\nu_1}
\newcommand{\nud}{\nu_2}
\newcommand{\nut}{\nu^a_3}

\newcommand{\ta}[2]{\tau^{#1}_{#2}}
\newcommand{\tta}[2]{\tilde{\tau}^{#1}_{#2}}
\newcommand{\la}[2]{\lambda^{#1}_{#2}}
\newcommand{\Ra}[2]{R^{#1}_{#2}}
\newcommand{\bRa}[2]{\bar{R}^{#1}_{#2}}
\newcommand{\La}[2]{L^{#1}_{#2}}
\newcommand{\Om}[2]{\Omega^{#1}_{#2}}
\newcommand{\tOm}[2]{\tilde{\Omega}^{#1}_{#2}}
\newcommand{\tL}[2]{\tilde{L}^{#1}_{#2}}

\newcommand{\Ups}[1]{\Upsilon_{#1}}

\newcommand{\Tt}{T^{a}_{3}}
\newcommand{\hH}{\hat{H}}
\newcommand{\bH}{\bar{H}}
\newcommand{\bJ}{\boldsymbol{J}}
\newcommand{\bR}{\boldsymbol{R}}
\newcommand{\bbR}{\boldsymbol{\bar{R}}}
\newcommand{\bL}{\boldsymbol{L}}
\newcommand{\bbL}{\boldsymbol{\bar{L}}}
\newcommand{\bla}{\boldsymbol{\lambda}}
\newcommand{\bbH}{\boldsymbol{\bar{H}}}
\newcommand{\tX}{\tilde{X}}

\newcommand{\DR}{\tilde{D}^{0\text{-}m}}
\newcommand{\DL}{D^{m\text{-}0}}
\newcommand{\DGS}{D^{0\text{-}0}}

\newcommand{\nbra}[1]{\bra*{#1}}
\newcommand{\nket}[1]{\ket*{#1}}
\newcommand{\rbra}{\bra*{\R}}
\newcommand{\rket}{\ket*{\R}}

\newcommand{\meq}{\mathrel{-}=}
\newcommand{\ssum}[2]{\sum_{\substack{#1 \\ #2}}}

\newcommand{\nv}[1]{n^{#1}_{\mathrm{v}}}
\newcommand{\no}[1]{n^{#1}_{\mathrm{o}}}
\newcommand{\NV}[1]{n^{#1}_{\mathrm{V}}}
\newcommand{\NO}[1]{n^{#1}_{\mathrm{O}}}

\newcommand{\half}{\frac{1}{2}}
\newcommand{\sixth}{\frac{1}{6}}

\newcommand{\mc}[3]{\multicolumn{#1}{#2}{#3}}
\newcommand{\mco}[1]{\mc{1}{c}{#1}}

\makeatletter
\def\lefteqno{\tagsleft@true}\def\righteqno{\tagsleft@false}
\makeatother

\author{Alexander C. Paul}
\author{Sarai D. Folkestad}
\author{Rolf H. Myhre}
\author{Henrik Koch}
\email{henrik.koch@sns.it}
\affiliation{Department of Chemistry, Norwegian University of Science and Technology,
             NTNU, 7491 Trondheim, Norway}
\alsoaffiliation{Scuola Normale Superiore, Piazza dei Cavaleri 7, 56126 Pisa, Italy}

\title[MLCC3]
{Oscillator strengths in the framework of equation of motion multilevel CC3 }

\abbreviations{CC, MLCC3}

\begin{document}

\begin{tocentry}
\end{tocentry}


\begin{abstract}
   We present an efficient implementation of the equation of motion oscillator strengths
   for the closed-shell multilevel coupled cluster singles and doubles
   with perturbative triples method (MLCC3) in the electronic structure program $e^T$.
   The orbital space is split into an active part
   treated with CC3 and an inactive part computed at the coupled cluster
   singles and doubles (CCSD) level of theory.
   Asymptotically,
   the CC3 contribution scales as $\mcl{O}(\NV{}\nv{3}\no{3})$
   floating-point operations (FLOP),
   where $\NV{}$ is the total number of virtual orbitals
   while $\nv{}$ and $\no{}$ are the number of active virtual
   and occupied orbitals, respectively.
   The CC3 contribution,
   thus, only scales linearly with the full system size and can become
   negligible compared to the cost of CCSD.
   We demonstrate the capabilities of our implementation
   by calculating the UV-VIS spectrum of azobenzene
   and a core excited state of betaine 30 with more than 1000 molecular orbitals.
\end{abstract}

\section{Introduction} \label{intro}

Coupled cluster theory is one of the most accurate models
when spectroscopic properties of small and medium sized molecules
are investigated.\cite{WavFuncRev,CCRespFunc, CCRespRev}
Due to its high accuracy and relatively feasible computational scaling
as $\mcl{O}(\NV{4}\NO{2})$,
CCSD is the most widely used variant of coupled cluster.
Despite its accuracy for valence excited states,
larger errors occur when considering core excited states or
double excitation dominated states.
\cite{comp_Xray_review,N2_core,XPS_benchmark,NOCI_core,glycine,CCSDR3}
Including triple excitations in the parametrization of the wave function 
improves the description of such states.
However, the computational cost and the memory requirement
increase to $\mcl{O}(\NV{5}\NO{3})$ and $\mcl{O}(\NV{3}\NO{3})$,
respectively for CCSDT.\cite{CCSDT,Bibel}
Approximating triples amplitudes with perturbation theory
can reduce the computational cost to $\mcl{O}(\NV{4}\NO{3})$
and the required memory to $\mcl{O}(\NV{2}\NO{2})$.

Triples corrections can be classified as iterative and noniterative models.
In noniterative models, corrections to the CCSD excitation energy are obtained
by expanding the excitation energy using many-body perturbation theory (MBPT).
The advantage of a noniterative approach
is that the triples correction is only computed once.
The disadvantage, 
however, 
is that transition moments cannot be easily defined.\cite{CCSDR3,size_int_trans}
The noniterative models include CCSDR(1a), CCSDR(1b) and CCSDR(3)
which are derived from the iterative methods CCSDT-1a, CCSDT-1b and CC3,
respectively.\cite{CCSDR3,CCSDTn,CC3_response,CC3_code,CC3_new}
Other noteworthy examples are CREOM-CCSD(T), EOMIP-CCSD$^*$
---developed specifically for ionized states---
and EOM-CCSD(T)(a)* which introduces
corrections to both the CCSD ground and the excited states.
\cite{CREOMptsd,ccsdstar_1,ccsdstar_2,ccsdptpa}

The best-known methods for including triples excitations iteratively are
CC3 and CCSDT-n.\cite{CC3_response, CCSDTn}
Both CCSDT-1 and CC3 scale asymptotically as $\mcl{O}(\NV{4}\NO{3})$,
but CC3 includes single excitations to infinite order leading to an improved
description of ground and excited states.\cite{CC3_response}
The advantage of iterative models is that they are more robust \cite{triples_properties}
and provide a consistent definition of other properties than the energy.
\cite{cc3_ee_benchmark}
However, that comes at the cost of iteratively converging equations scaling
as $\mcl{O}(\NV{4}\NO{3})$.
Nevertheless, with current implementations systems of around 400 basis functions
can be routinely treated at the CC3 level.\cite{CC3_new}

Due to the success of coupled cluster theory,
schemes have been developed to reduce the scaling while keeping the accuracy.
Pulay and S\ae{}b\o{} advocated the use of localized molecular orbitals (LMOs),
for a compact description of electronic correlation
in M\o{}ller-Plesset (MP) perturbation theory and configuration interaction
singles and doubles(CISD).
\cite{Local_corr_1,Local_corr_2,Local_CI,Local_MP_1,Local_MP_2}
They used Boys localization for the occupied molecular orbitals
and projected atomic orbitals (PAOs) for the virtual space,
and reduced the scaling by neglecting the correlation between distant pairs of localized orbitals.
\cite{Local_corr_2}
Werner and Sch\"utz then extended this model to coupled cluster theory
with and without a noniterative triples correction.
\cite{Local_CC,Local_CCSDpT_1,Local_CCSDpT_2}
Domain based local pair-natural orbital coupled cluster (DLPNO-CC)
methods are also related to this approach.\cite{Neese_LPNO,Neese_LPNO_CC}
The DLPNO-CC approach has recently been extended to
CCSD(T) and also CC3 which was used to calculate the first electronic excited state
of a system with more than 1300 basis functions.\cite{DLPNO_CCSDpT,DLPNO_CC3}
Reducing the size of the active space based on a distance criterium is certainly
successful for ground state properties.
For the description of excitation energies and other excited state properties,
however,
distance measures do not work as well
as more diffuse orbitals become more important.
\cite{Local_EE_CCSD,Local_CC2,Local_CC_prop,Local_laplace_CC2,
PNO_CC2,Towards_local_EE,Accuracy_LocalEOM}
Therefore,
larger active spaces have to be employed in these calculations
and different orbital spaces are used for the ground and excited states.
\cite{Local_EE_CCSD,Local_CC2}

Multilevel and embedding methods treat different regions of a system
with different levels of theory.
The idea of obtaining an accurate description of a large molecular system
by coupling the contributions of its subsystems is exploited in
QM/MM approaches,\cite{QMMM_1,QMMM_2,QMMM_3,QMMM_review_1,QMMM_review_2,PElib}
frozen density embedding,\cite{FDE_1,FDE_2}
subsystem DFT,\cite{SubsystemDFT_1,SubsystemDFT_2}
and the ONIOM, IMOMO and LMOMO methods.\cite{IMOMO,ONIOM,LMOMO}
Another method related to multilevel coupled cluster (MLCC)
was developed by Oliphant and Adamowicz using CCSD for multireference
systems by including selected triple and quadruple substitutions.
\cite{ML_MRCC_1,ML_MRCC_2,ML_MRCC_3}
This scheme was adapted by K\"{o}hn and Olsen to include higher order substitutions
at reduced cost.\cite{Active_CC,Active_CC_TQ}

In multilevel coupled cluster (MLCC) one CC wave function
is used for the full system but different parts of the system
are described with different level of truncation.\cite{MLCCSD_ECC2,MLCC}
Considerable savings are achieved by applying the higher order excitation operators
in a smaller (active) subset of the orbitals.\cite{MLCC3}
The active orbital space can be selected using localized orbitals
--- such as Cholesky orbitals\cite{Cholesky_Orbs} and projected atomic orbitals (PAOs)
\cite{Local_corr_2}
--- or state-selective approaches ---
such as the correlated natural transition orbitals (CNTOs).\cite{MLCC_CNTO_CVS}
As MLCC is designed for intensive properties, excitation energies
or oscillator strengths are accurately reproduced
if an appropriate active space is chosen.\cite{MLCC_CNTO_CVS,MLCC3,MLCC_CVS,MLCC_EOM}
While state-selective approaches are preferred to keep
the active space as compact as possible,
they are less suited for transition properties
especially between excited states,
as a consistent active space is needed for all excited states.\cite{DLPNO_CC3}
Localized orbitals are only suitable in the cases where the target property 
is localized in a smaller region of the molecule.

In this paper we report the extension of the MLCC3 method to compute
oscillator strengths with CC3 quality but at significantly reduced cost.
Employing core-valence separation (CVS),
oscillator strengths are also available for core excited states.\cite{CVS,CVS_ADC,CVS_CC}
This allows us to tackle excited states and oscillator strengths of systems
with more than 1000 basis functions.

\section{Theory} \label{Theory}

In this section,
we will introduce the closed shell MLCC3 model within the equation of motion (EOM) formalism.
For a more detailed derivation we refer to Refs \citenum{CC3_response,CC3_new}.
Consider the general cluster operator
\begin{equation}
   T = \sum_{\mu} \tau_\mu X_\mu,
\end{equation}
where $X_\mu$ is an excitation operator that converts the reference
determinant, $\rket$,
into the excited determinant, $\nket{\mu}$,
and $\tau_\mu$ is the corresponding amplitude.
In MLCC3 with two levels,
namely CCSD and CC3,
the cluster operator assumes the form
\begin{equation}
   T = T_1 + T_2 + \Tt
\end{equation}
with
\begin{equation}
\begin{split}
   T_1 &= \ssum{A}{I} \ta{A}{I} E_{AI} \\
   T_2 &= \half \ssum{AB}{IJ} \ta{AB}{IJ} E_{AI} E_{BJ} \\
   \Tt &= \sixth \ssum{abc}{ijk} \ta{abc}{ijk} E_{ai} E_{bj} E_{ck}
\end{split}
\label{eq:T_op}
\end{equation}
where $E_{AI}$ and $E_{ai}$ are singlet excitation operators.
While the operators $T_1$ and $T_2$ excite on the full orbital space
indicated by capitalized indices,
the triples cluster operator $\Tt$ only excites in the active orbital space
denoted by lower case indices.
We use the standard notation
where the indices $i$, $j$, $k\dots$ refer to occupied,
$a$, $b$, $c\dots$ to virtual,
and $p$, $q$, $r\dots$ to general active orbitals.
The CC wave function is defined as
\begin{equation}
   \nket{\CC} = \exp(T) \rket
\end{equation}
and we introduce the similarity transformed Hamiltonian
\begin{equation}
   \bH = \exp(-T) \hH \exp(T)
\end{equation}
where
\begin{equation}
   \hH = \sum_{pq} h_{pq} E_{pq}
     + \half \sum_{pqrs} (pq|rs) (E_{pq} E_{rs} - E_{ps} \del{qr}) + h_{nuc}
\end{equation}
is the electronic Hamiltonian.
To obtain the cluster amplitudes a set of biorthogonal determinants
\begin{equation}
   \{\nbra{\mu}\} = \{\nbra{\mu_1}\} \oplus \{\nbra{\mu_2}\} \oplus \{\nbra{\mu_3^a}\}
   \label{eq:bo_space}
\end{equation}
is defined,
where the triply excited determinants, $\mu_3^a$,
are restricted to the active space.
These determinants are generated using the contravariant excitation operator,
$\tX_\mu$, such that,
\begin{equation}
   \braket{\mu}{\nu} = \rbra \tX_\mu X_\nu \rket = \del{\mu\nu}.
   \label{eq:Omega}
\end{equation}
The coupled cluster energy, $\ECC$,
and the cluster amplitudes are then obtained by projection
onto the reference determinant and the set of excited determinants,
respectively,\cite{Bibel}
\begin{align}
   \label{eq:energy}
   \ECC &= \rbra \bH \rket
   \\
   \Om{}{\mu} &= \nbra{\mu} \bH \rket = 0.
   \label{eq:amplitudes}
\end{align}
To obtain compact equations we incorporate the effect of the singles
cluster operator into the Hamiltonian and obtain the so-called
$T_1$-transformed Hamiltonian,
\begin{equation}
   H = \exp(-T_1) \hH \exp(T_1).
\end{equation}
In analogy to MBPT,
the $T_1$-transformed Hamiltonian is split into an
effective one-particle operator and a fluctuation potential.
\begin{equation}
   H = F + U
\end{equation}
In CC3 the double excitation amplitudes and the fluctuation potential are
treated as first order in the perturbation while the triples amplitudes
are considered second order.
The single excitation amplitudes are included as zeroth order parameters,
as they have a special role as relaxation parameters.\cite{CC3_response,CC3_code}
Inserting eq \eqref{eq:T_op} and eq \eqref{eq:bo_space} into eq \eqref{eq:Omega} and neglecting
all terms of third and higher order in the perturbation,
we obtain the MLCC3 ground state equations,
\begin{align}
   \label{eq:OM_CC3_S}
   \Om{}{\mus} &= \nbra{\mus} H + [H,T_2] + [H,\Tt] \rket \\
   \label{eq:OM_CC3_D}
   \Om{}{\mud} &= \nbra{\mud} H + [H,T_2] + [[H,T_2],T_2] + [H,\Tt] \rket \\
   \label{eq:OM_CC3_T}
   \Om{}{\mut} &= \nbra{\mut} [H,T_2] + [F,\Tt] \rket.
\end{align}
The Fock matrix is not necessarily diagonal in the local orbital basis,
but it can be block-diagonalized within the active orbital space,
such that the off-diagonal elements do not contribute to the triples amplitudes.
Therefore,
the triples amplitudes can be expressed in terms of the doubles amplitudes
\begin{equation}
   \ta{abc}{ijk} = -\frac{1}{\eps^{abc}_{ijk}} \nbra{\mut} [H,T_2] \rket,
\end{equation}
where $\eps^{abc}_{ijk}$ are the orbital energy differences
\begin{equation}
   \eps^{abc}_{ijk} = \eps_{a}+\eps_{b}+\eps_{c}-\eps_{i}-\eps_{j}-\eps_{k}.
\end{equation}

In equation of motion coupled cluster (EOM-CC)
start out from the matrix representation of the similarity transformed Hamiltonian,
\begin{equation}
   \label{eq:H_mat_el}
   \bbH =
   \begin{pmatrix}
      \rbra \bH \rket      & \rbra \bH \nket{\nu} \\
      \nbra{\mu} \bH \rket & \nbra{\mu} \bH \nket{\nu}
   \end{pmatrix}.
\end{equation}
If the CC ground state equations, eq \eqref{eq:Omega},
are converged,
the similarity transformed Hamiltonian can be written as,
\begin{equation}
   \label{eq:H_mat}
   \bbH =
   \begin{pmatrix}
      0      & \bm{\eta}^T \\
      \bm{0} & \bJ
   \end{pmatrix}
   +
   \ECC \bm{I},
\end{equation}
where $\eta_\nu = \rbra [\bH,X_\nu] \rket$
and $\bJ$ is the so-called Jacobian with matrix elements
$\nbra{\mu} [\bH,X_\nu] \rket$.
The eigenvectors of $\bH$ are the EOM states and the corresponding
eigenvalues the energies of these states.
As the similarity transformed Hamiltonian is non-symmetric,
the left and right eigenvectors are not hermitian conjugates,
but they are biorthonormal.\cite{Bibel}
\begin{equation}
   \bbH\bbR_m = E_m\bbR_m
   \quad
   \bbL^T_m\bbH = E_m\bbL^T_m
   \quad
   \bbL^T_m\bbR_n = \del{mn}
\end{equation}
From the biorthogonality of the EOM states and the structure of the Hamiltonian matrix,
we obtain the left and the right ground state,
\begin{equation}
   \label{eq:GS_vec}
   \bbL_0 =
   \begin{pmatrix}
      1      \\
      \bla
   \end{pmatrix}
   \quad
   \bbR_0 =
   \begin{pmatrix}
      1      \\
      \bm{0}
   \end{pmatrix},
\end{equation}
and the left and right excited states,\cite{CC3_new}
\begin{equation}
   \label{eq:ES_vec}
   \bbL_m =
   \begin{pmatrix}
      0      \\
      \bL_m
   \end{pmatrix}
   \quad
   \bbR_m =
   \begin{pmatrix}
      -\bla \bR_m \\
      \bR_m
   \end{pmatrix}.
\end{equation}
The parameters $\bla$
are determined from
\begin{equation}
   \bla^T \bJ = - \bm{\eta},
\end{equation}
while the parameters of the excited states are determined as eigenvectors of the Jacobian,
$\bJ$.
The MLCC3 Jacobian is given by\cite{MLCC3}
\begin{equation}\label{eq:jcc3}
\footnotesize
   \bJ^{MLCC3} =
   \begin{pmatrix}
      \nbra{\mus}[H + [H,T_2],X_{\nus}]\rket &
      \nbra{\mus}[H,X_{\nud}]\rket &
      \nbra{\mus}[H,X_{\nut}]\rket
      \\
      \nbra{\mud}[H + [H,T_2+\Tt],X_{\nus}]\rket &
      \nbra{\mud}[H + [H,T_2],X_{\nud}]\rket &
      \nbra{\mud}[H,X_{\nut}]\rket
      \\
      \nbra{\mut}[H + [H,T_2], X_{\nus}]\rket &
      \nbra{\mut}[H,X_{\nud}]\rket &
      \nbra{\mut}[F,X_{\nut}]\rket
   \end{pmatrix}.
\end{equation}
The vectors in eq \eqref{eq:GS_vec} and eq \eqref{eq:ES_vec} correspond to operators
which generate the EOM states from the Hartree-Fock determinant.
\begin{align}
   \nbra{\tCC} &= \rbra \Big(1 + \sum_\mu \la{}{\mu} \tX_\mu \Big) \exp(-T) \\
   \nket{\CC} &= \exp(T) \rket \\[0.2eM]
   \nbra{m} &= \rbra \sum_\mu \La{}{\mu} \tX_\mu \exp(-T) \\
   \nket{m} &= \Big(\sum_\mu \Ra{}{\mu} X_\mu - \sum_\mu \Ra{}{\mu} \la{}{\mu}\Big)
   \exp(T) \rket
\end{align}
Once the ground and excited states are determined,
left and right transition moments can be obtained in terms of
left ($\DL$) and right ($\DR$) transition densities.
\cite{EOMCC,EOM_densities,EOM_gradient}
\begin{align}
   \nbra{\tCC} A \nket{m} &= \sum_{pq} \DR_{pq} A_{pq} \\
   \nbra{m} A \nket{\CC} &= \sum_{pq} \DL_{pq} A_{pq}
\end{align}
Here,
$A$ is a general one-electron operator $A=\sum_{pq} A_{pq} E_{pq}$.

To obtain accurate excitation energies and transition dipole moments,
the selection of the active orbital space is crucial.
In this paper two approaches are chosen to partition the orbital space.
For the cheaper strategy Cholesky orbitals are used for the occupied space.
To obtain these orbitals the Hartree-Fock density is Cholesky decomposed
using the AOs of the active atoms as pivoting elements.
\cite{Cholesky_Orbs,Cholesky_subsystems}
\begin{equation}
   D_{\alpha\beta} = \sum_J C^a_{\alpha J} C^a_{\beta J} + \Delta D_{\alpha\beta}
\end{equation}
The decomposition is stopped when the size of all active diagonal elements
is below a given threshold and the coefficients are simply the elements
of the Cholesky vectors $C_{\alpha J}$.
The inactive orbitals are then obtained by decomposing the remaining part of the density,
$\bm{\Delta D}$.
Projected atomic orbitals have been shown to give a good description of the
virtual space for solvated systems,
but also adenosine. \cite{MLCC_EOM,MLCC_MLHF,MLCC_Embedding}

The construction of correlated natural transition orbitals is more costly
as they are obtained from excitation vectors of a coupled cluster calculation.
In MLCC3 we use CNTOs constructed from CCSD excited states
to get a compact description of the excited states.
The CNTOs are generated by diagonalizing two matrices,
denoted by $\bm{M}$ and $\bm{N}$,
defined as
\begin{align}
   \label{eq:M}
   M_{ij} = \sum_a R^a_i R^a_j + \frac{1}{2}\ssum{ab}{k} (1+\del{ai,bk}\del{ij}) \Ra{ab}{ik} \Ra{ab}{jk}\\
   \label{eq:N}
   N_{ab} = \sum_i R^a_i R^b_i + \frac{1}{2}\ssum{c}{ij} (1+\del{ai,cj}\del{ab}) \Ra{ac}{ij} \Ra{bc}{ij}.
\end{align}
The eigenvectors of $\bm{M}$ and $\bm{N}$ correspond to the CNTO transformation
matrices for the occupied and virtual CNTOs, respectively.
The CNTOs whose eigenvalues sum up to a certain cutoff are chosen as active space
\begin{align}
   \label{e:cnto_threshold_o}
   1 - \xi_M &< \sum_o \lambda^M_o \\
   \label{e:cnto_threshold_v}
   1 - \xi_N &< \sum_v \lambda^N_v
\end{align}
where $\lambda^M_o$ and $\lambda^N_v$ are the eigenvalues of $\bm{M}$ and $\bm{N}$.
To obtain the most compact basis, separate CNTO bases for each excited state
would be preferable.
However,
due to the non-orthogonality of the orbitals,
subsequent calculation of transition moments between excited states would be complicated.
Therefore,
we choose a state averaged approach,
\begin{equation}
   \bm{M} = \frac{1}{n_{ES}} \sum^{n_{ES}}_i \bm{M}_i, \quad
   \bm{N} = \frac{1}{n_{ES}} \sum^{n_{ES}}_i \bm{N}_i,
\end{equation}
where $\bm{M}_i$ and $\bm{N}_i$ are constructed according to eq \eqref{eq:M}
and \eqref{eq:N} for the $i$-th excited state
and $n_{ES}$ is the number of excited states included in the matrices.

\section{Implementation}\label{Implementation}

The closed shell MLCC3 ground and excited states as well as
EOM transition properties have been implemented in the $e^T$ program package.\cite{eT}
One of the advantages of MLCC3 compared to other reduced cost methods is
that only the space,
in which the triples amplitudes are defined,
is restricted.
Therefore,
we can split the occupied and virtual orbitals into active
and inactive subsets,
and use almost identical code for MLCC3 as for full CC3.
The algorithms employed to calculate closed shell CC3 properties
in $e^T$ have been detailed in Ref. \citenum{CC3_new}
and only a short summary will be given in this paper.
The ground state residual, $\bm{\Omega}$,
and the transformations of a trial vector with the Jacobian
are computed in a restricted loop over the occupied indices $i\geq j\geq k$.
An $\nv{3}$-block of triples amplitudes is constructed for a given
set of indices $\{i,j,k\}$.
Using this structure, the permutational symmetry of the triples amplitudes
can be exploited,
while utilizing efficient matrix multiplication routines
for the contractions of the block of virtual orbitals.
\cite{CCSDpT_Lee,Triples_alg1,Triples_alg2}
By reformulating the equations in terms of contravariant triples amplitudes
\begin{equation}\label{eq:contra_3}
   \tta{abc}{ijk} = 4\ta{abc}{ijk} - 2\ta{bac}{jik} - 2\ta{cba}{kji}
                  - 2\ta{acb}{ikj} + \ta{cab}{kij} + \ta{bca}{jki},
\end{equation}
and residuals,
$\tOm{}{}$,
the number of memory-bound reordering operations is reduced.
After all contributions to the contravariant residual are collected
it is converted back to the covariant form,
using the relations
\begin{align}
   \tOm{A}{I} &= \Om{A}{I} \\
   \tOm{AB}{IJ} &= 2\Om{AB}{IJ} - \Om{BA}{IJ}, \quad
   \Om{AB}{IJ} = \frac{1}{3}(2\tOm{AB}{IJ} + \tOm{BA}{IJ}).
\end{align}
As in CC3,
the $\ta{}{3}$ amplitudes are defined in terms of the $\ta{}{2}$ amplitudes
\begin{equation}
   \label{eq:tau3}
   \ta{abc}{ijk} = -(\eps^{abc}_{ijk})^{-1} P^{abc}_{ijk}
                     \Big(\sum_D \ta{aD}{ij} g_{bDck} -
                     \sum_L \ta{ab}{iL} g_{Ljck}\Big).
\end{equation}
However,
because the triples determinants are restricted to the active space
only the summation indices in the expression for $\ta{}{3}$ are over the full space.
Here,
$P^{abc}_{ijk}$ is a permutation operator creating a sum of all unique permutations
of the index pairs $ai, bj, ck$,
and $g_{pqrs}$ are two-electron integrals in the $T_1$-trasformed basis.\cite{Bibel}
From eq \eqref{eq:tau3} it is evident that the most memory efficient implementation
will make use of two separate arrays for $\ta{ab}{iL}$ and $\ta{aD}{ij}$.
Similarly,
two vectors are needed for the doubles part of the ground state residual
because one index originates from a $T_1$-transformed two-electron integral, 
$g_{pqrs}$,
\begin{align}
   \tOm{ab}{iL} &= \ssum{c}{jk} \tta{abc}{ijk} g_{jLkc} \\
   \tOm{aD}{ij} &= \ssum{bc}{j} \tta{abc}{ijk} g_{Dbkc},
\end{align}
Therefore,
the memory requirement and the computational cost of the triples contributions
scale linearly with the full size of the system,
and the overall asymptotic scaling for constructing the ground state residual is
$4\NV{}\nv{3}\no{3}$ floating point operations (FLOP).

The triples amplitudes of the right excitation vector can be expressed as
\begin{equation}
   \label{eq:R3}
   \Ra{abc}{ijk} = -\frac{1}{\eps^{abc}_{ijk} - \omega} P^{abc}_{ijk} \Big(
   \sum_D \bRa{aD}{ij} g_{bDck}   - \sum_L \bRa{ab}{iL} g_{Ljck}
   + \sum_D \ta{aD}{ij} \Ups{bDck} - \sum_L \ta{ab}{iL} \Ups{Ljck} \Big)
\end{equation}
where $\Ups{bDck}$ and $\Ups{Ljck}$ are treated as one-index transformed integrals
\begin{align}
   \label{eq:ups_v}
   &\Ups{bDck} = \sum_E R^E_k g_{bDcE}
               - \sum_M \big(R^b_M g_{MDck} + R^c_m g_{bDMk}\big) \\
   \label{eq:ups_o}
   &\Ups{Ljck} = \sum_E \big(R^E_j g_{LEck} + R^E_k g_{LjcE}\big)
               - \sum_M R^c_M g_{LjMk}.
\end{align}
and $\bRa{ab}{ij} = (1+\del{ai,bj})\Ra{ab}{ij}$.\cite{Bibel}
From eq \eqref{eq:R3} can be seen that the construction of $R_3$ is
twice as expensive as the construction of $\tau_3$.
For the Jacobian transformation the same terms have to be computed
as for the ground state residual,
but $R_3$ is contracted instead of $\tau_3$.
%
%
Additionally,
the $\tau_3$ amplitudes are required for a single term
leading to an overall asymptotic scaling of $8\NV{}\nv{3}\no{3}$ FLOP.
It should be noted that the construction of $\Ups{bDck}$
scales quadratically with the full system size.
However,
this term will not be significant compared to the
other terms in the Jacobian transformation.

The transpose Jacobian transformation also scales with $8\NV{}\nv{3}\no{3}$ FLOP,
as the $L_3$ and $\tau_3$ amplitudes need to be constructed and two contractions,
each scaling as $2\NV{}\nv{3}\no{3}$ FLOP, are needed.
%
%
The final contractions contributing to the singles part of the transformed vector
contains terms that scale quadratically with the full size of the system.
However,
these terms scale at most as $2\NV{}\NO{}\nv{2}\no{}$ FLOP and are therefore
negligible compared to full CCSD.

To obtain core excited states core-valence separation is employed,
where all non-zero elements of both the trial vector and the transformed vector
need to contain at least one index corresponding to a core orbital.
\cite{CVS_CC,MLCC_CVS,CC3_new}
Therefore,
in this implementation of the Jacobian transformations, 
we skip iterations in the loop over $i,j,k$ 
if all indices correspond to valence orbitals.
This reduces the scaling for both Jacobian transformations to $8\NV{}\nv{3}\no{2}$.

As in the full CC3 code the EOM transition densities are constructed in
a loop over the occupied indices and another loop over the virtual indices.
We calculate all contributions to the density in a loop over the occupied indices,
except for one contribution to the occupied-occupied block of the density
which cannot be efficiently calculated in a loop over $i,j,k$.
\begin{equation}
   \label{eq:DL_oo}
   \DL_{kl} \meq \frac{1}{2} \ssum{abc}{ij} \tL{abc}{ijl} \ta{abc}{ijk}
\end{equation}
As shown in eq \eqref{eq:DL_oo} for the occupied-occupied block of the left transition density,
the triples amplitudes that are contracted differ in the occupied indices.
Therefore,
a triples loop over the virtual indices has to be used, 
in order to exploit the permutational symmetry of the triples amplitudes.
This leads to an increase in contractions scaling as $2\NV{}\nv{3}\no{3}$ FLOP.
However,
the triples amplitudes have to be reconstructed for the loop over $a,b,c$
which also leads to a larger prefactor in the scaling.
While the contractions inside the triple loops scale linearly with
the full system size,
there exists one term in the right transition density, $\DR$,
that requires storing a subblock of $\ta{}{2}$ scaling as $\NV{}\NO{}\nv{}\no{}$
in memory.
This is, however, not an issue as CCSD is used as lower level method
where the full $\ta{}{2}$ array scaling as $\NV{2}\NO{2}$ needs to be kept in memory.

Because the triples amplitudes have to be calculated twice
the overall scaling to construct a single $\DL$ amounts to $10\NV{}\nv{3}\no{3}$ FLOP.
The construction of a single $\DR$ totals $16\NV{}\nv{3}\no{3}$ FLOP,
as the $\Ra{}{3}$ amplitudes are twice as expensive as the $\La{}{3}$,
and also the $\ta{}{3}$ and $\la{}{3}$ amplitudes are required.
For transition moments from the ground state,
these densities only need to be computed once per state,
compared to the iterative cost (per state) for the Jacobian transformations.

\section{Results and Discussion}\label{Applications}
With the MLCC3 method, we can obtain
excitation energies and oscillator strengths
of CC3 quality at significantly reduced cost.
We compare the MLCC3 results for oxygen core excitations of guanine to the CC3
results.
The scaling with the size of the inactive space is shown for formaldehyde
with up to six explicit water molecules.
To show the capabilities of the method,
the UV/VIS spectrum of azobenzene
and a core excited state of betaine 30 with more than 1000 molecular orbitals
are reported.

\subsection{Guanine}
%
A single core excited state of the oxygen atom of guanine is calculated
with aug-cc-pCVDZ basis set on the oxygen atom and aug-cc-pVDZ on the remaining atoms
using two Intel Xeon-Gold 6138 with 40 threads in total.
\cite{aug_cc_pVXZ,cc_pCVXZ,cc_pVXZ}
%
The results and timings per iteration are summarized in Table \ref{t:Gua_1state}
for selected active spaces.
The number of virtual orbitals is chosen to be 10 times larger than the number
of occupied orbitals.
Already with an active space comprising 10 occupied orbitals the excitation
energies improve by 2\,eV compared to CCSD and the difference to CC3 is only 0.4\,eV.
Increasing the active space to 15 occupied orbitals the deviation from the CC3 results
is below 0.2\,eV.
For 15 occupied orbitals
the error of MLCC3 is below the expected error of CC3 for oxygen core excitations.
\begin{table}
\centering
\caption{Timings in seconds to compute a core excited state          from the oxygen atom of guanine
         at the CCSD and MLCC3 level with several active spaces.
         Timings are given, averaged over the number of iterations when solving for
         $\bm{\tau}$, $\bm{\lambda}$, $\bm{R}$ and $\bm{L}$.
         Additionally, timings to construct the ground state density, $\bm{\DGS}$,
         left transition density, $\bm{\DL}$, and right transition density, $\bm{\DR}$,
         are reported.
         Note that the MLCC3 and CC3 timings only comprise the triples part.}
\label{t:Gua_1state}
\begin{tabular}{l SS SSSSS}
   \toprule
   & \mco{CCSD} & \mc{5}{c}{MLCC3} & \mco{CC3} \\
   \cmidrule(lr){2-2}\cmidrule(lr){8-8}\cmidrule(lr){3-7}
   $\no{}/\nv{}$ & & \mco{$10/100$} & \mco{$13/130$} & \mco{$15/150$} & \mco{$18/180$} & \mco{$20/200$} &  \\
   \midrule
   $\omega$ [eV] & 535.91 & 533.90 & 533.76 & 533.69 & 533.61 & 533.58 & 533.51 \\
   $f\times 100$ & 3.26 & 2.42 & 2.31 & 2.26 & 2.20 & 2.18 & 2.12 \\
   \midrule
   $\bm{\tau}$    & 15.49 & 3.03 & 10.72 & 24.93 & 70.31 & 125.74 & 2220.55 \\
   $\bm{\lambda}$ & 25.78 & 5.40 & 21.49 & 46.08 & 130.22 & 238.02 & 4157.37 \\
   $\bm{R}$       & 24.48 & 1.72 & 4.82 & 9.22 & 20.90 & 33.13 & 301.98 \\
   $\bm{L}$       & 23.62 & 1.86 & 5.27 & 9.83 & 22.59 & 36.58 & 317.90 \\
   $\bm{\DGS}$    & 0.59 & 5.80 & 25.76 & 66.43 & 175.06 & 312.30 & 5147.35 \\
   $\bm{\DL}$     & 0.21 & 3.68 & 14.76 & 33.33 & 95.24 & 171.81 & 2340.44 \\
   $\bm{\DR}$     & 0.71 & 7.29 & 29.97 & 65.50 & 182.68 & 320.64 & 4638.42 \\
   \bottomrule
\end{tabular}
\end{table}
\begin{table}
\centering
\caption{Speed up of MLCC3 compared to CC3 calculated according
         to equations \ref{e:speedup_gs} and \ref{e:speedup_es}.
         The first part shows the speed up for terms that scale asymptotically
         as $\mcl{O}(\NV{}\nv{3}\no{3})$ while the second part summarizes the
         speed up for terms with a cost of $\mcl{O}(\NV{}\nv{3}\no{2})$.}
\label{t:Gua_speedup_1}
\begin{tabular}{l c
                S[table-number-alignment = center]
                S[table-number-alignment = center]
                S[table-number-alignment = center]
                S[table-number-alignment = center]
                S[table-number-alignment = center]}
   \toprule
   $\no{}$ && \mco{10} & \mco{13} & \mco{15} & \mco{18} & \mco{20} \\
   $\nv{}$ && \mco{100} & \mco{130} & \mco{150} & \mco{180} & \mco{200} \\
   \midrule
   $\bm{\tau}$      &&  732.9 & 207.1 & 89.1 & 31.6 & 17.7 \\
   $\bm{\lambda}$   &&  769.9 & 193.5 & 90.2 & 31.9 & 17.5 \\
   $\bm{\DGS}$      &&  887.5 & 199.8 & 77.5 & 29.4 & 16.5 \\
   $\bm{\DL}$       &&  636.0 & 158.6 & 70.2 & 24.6 & 13.6 \\
   $\bm{\DR}$       &&  636.3 & 154.8 & 70.8 & 25.4 & 14.5 \\
   \cmidrule{1-1}
   $S^{GS}_{theo}$  && 1079.1 & 223.6 & 94.7 & 31.7 & 16.9 \\
   \midrule
   $\bm{R}$         &&  175.6 & 62.7 & 32.6 & 14.5 & 9.1 \\
   $\bm{L}$         &&  170.9 & 60.3 & 32.3 & 14.1 & 8.7 \\
   \cmidrule{1-1}
   $S^{ES}_{theo}$  &&  276.7 & 74.5 & 36.4 & 14.6 & 8.7 \\
   \bottomrule
\end{tabular}
\end{table}
For the smallest active space in Table \ref{t:Gua_1state} the cost per iteration
is much smaller than the CCSD timings.
The CC3 contribution dominates inly in the construction of the densities,
because CCSD densities scale as $\mcl{O}(\NV{3}\NO{2})$ in contrast
to $\mcl{O}(\NV{}\nv{3}\no{3})$ for MLCC3 densities.
Considering active spaces with 13 and 15 occupied orbitals,
the time spent in the MLCC3 part of the code is almost identical to the
time in the CCSD code.
The excited states are significantly cheaper with MLCC3,
as CVS is implemented by skipping iterations in the $i$, $j$, $k$ loop,
effectively reducing the scaling to $\mcl{O}(\NV{}\nv{3}\no{2})$.

In Table \ref{t:Gua_speedup_1} we report speed up compared to CC3.
For terms scaling as $\mcl{O}(\NV{}\nv{3}\no{3})$ the speed up is calculated as,
\begin{equation}
   \label{e:speedup_gs}
   S^{GS} = \frac{t^{it}_{CC3}}{t^{it}_{MLCC3}} \qquad
   S^{GS}_{theo} = \frac{(\NO{}\times \NV{})^3}{(\no{}\times \nv{})^3},
\end{equation}
while for core excited states the reduction in the scaling is given by,
\begin{equation}
   \label{e:speedup_es}
   S^{ES} = \frac{t^{it}_{CC3}}{t^{it}_{MLCC3}} \qquad
   S^{ES}_{theo} = \frac{\NO{2}\times \NV{3}}{\no{2}\times \nv{3}}.
\end{equation}
%
It should be noted that only the dominating terms are included in this estimate,
but terms with a lower scaling can be significant,
especially for small active spaces.
With an active space of 15 occupied orbitals a speed up of about 90 can be reached,
while the deviation from the CC3 results is below 0.2\,eV.

As we pursue a state-averaged approach in the determination of the active space,
the performance is expected to deteriorate somewhat when more states are considered.
Four core excited states of the oxygen atom of guanine are calculated
with aug-cc-pCVDZ basis set on the oxygen atom and aug-cc-pVDZ on the remaining atoms
\cite{aug_cc_pVXZ,cc_pCVXZ,cc_pVXZ}.
The calculations were performed on two Intel Xeon E5-2699 v4 processors using 40 threads,
so the timings are not directly comparable to Table \ref{t:Gua_1state}.

Instead of specifying active spaces explicitly,
we chose to use the CNTO threshold as defined
in eq \ref{e:cnto_threshold_o} and \ref{e:cnto_threshold_v}.
For a more direct comparison the results of calculations
performed as above are tabulated in the SI (Table \ref{si-t:Gua_4_systematic}).
Both the thresholds for the occupied and virtual orbital space
are reduced from $10^{-1}$ to $10^{-6}$ while keeping both thresholds at the same magnitude.
The size of the active spaces and the full size of the system are summarized
in Table \ref{t:cnto_space_4}.
By using the thresholds,
the ratio between active virtual orbitals and active occupied orbitals
reduces to approximately 7.

\begin{figure}[ht]
   \centering
   \begin{subfigure}{0.56\textwidth}
      \includegraphics[width=\textwidth]{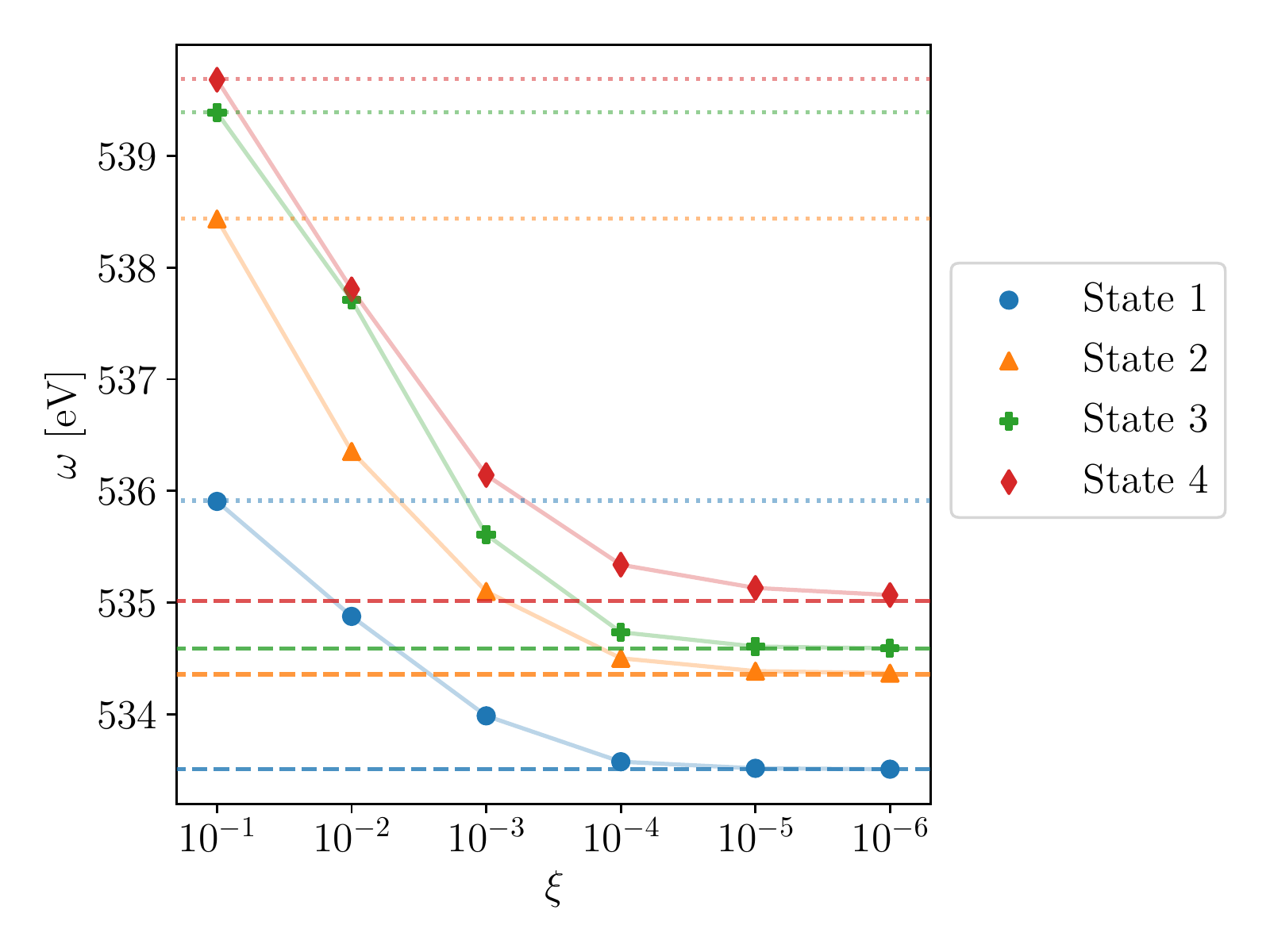}
   \end{subfigure}
   \begin{subfigure}{0.43\textwidth}
      \includegraphics[width=\textwidth]{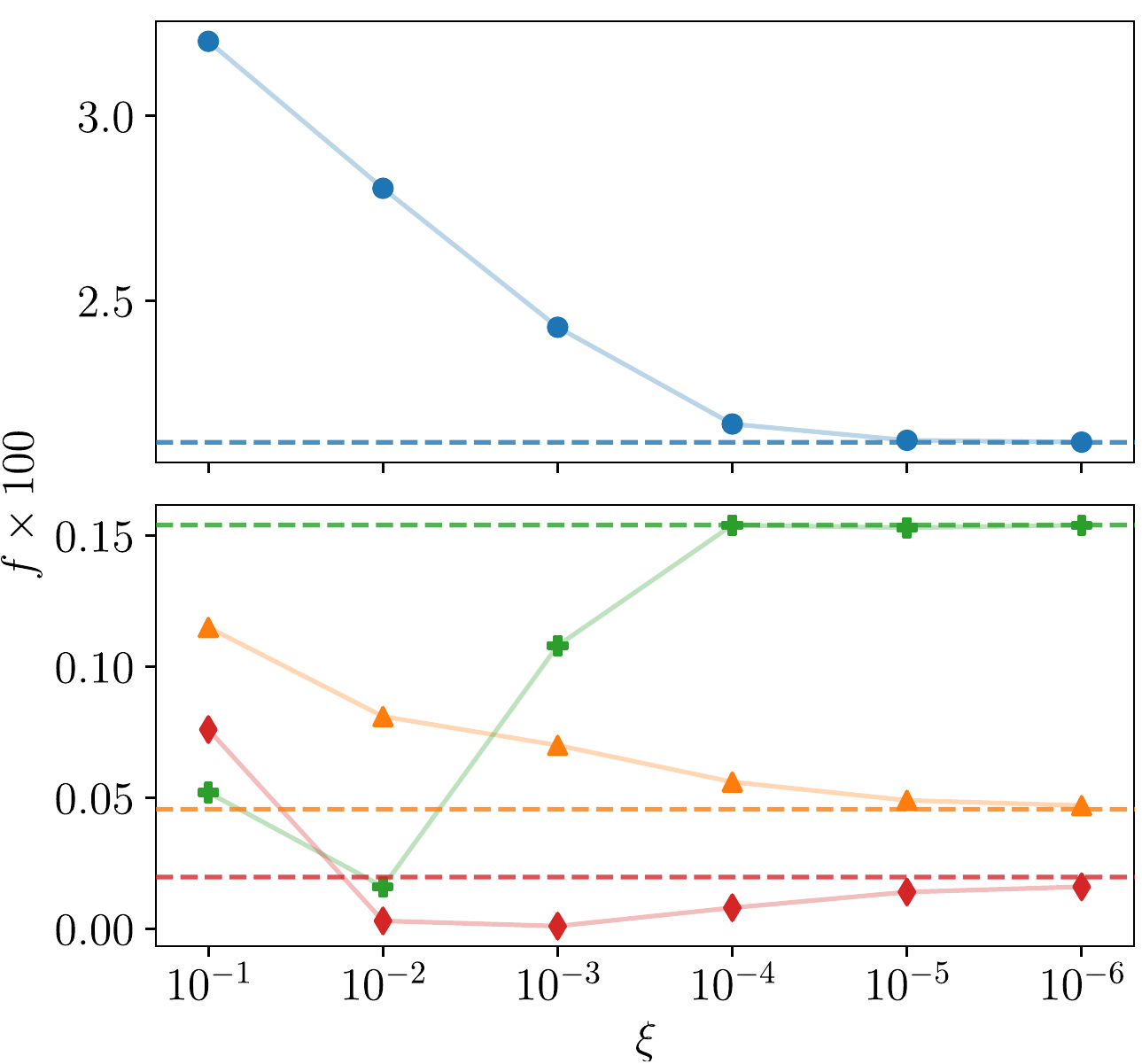}
   \end{subfigure}
   \caption{Convergence of the first four core excitation energies (left)
            and oscillator strengths (right) of guanine with CNTO threshold.
            Dashed lines are the CC3 results and dotted lines denote the CCSD values.}
   \label{fig:cntoG}
\end{figure}

The excitation energies, $\omega$,
and oscillator strengths, $f$,
are reported in Table \ref{t:Gua_4states}.
For a threshold of $10^{-1}$ the occupied orbital space
consists only of a single orbital,
such that the triples amplitudes are zero by definition.
The results for this threshold are always identical to CCSD.
%
\begin{table}
\centering
\caption{Number of occupied and virtual orbitals in the active space
         for guanine for various CNTO thresholds.
         The CNTOs have been constructed from four core excited states obtained
         at the CCSD level of theory.}
\label{t:cnto_space_4}
\begin{tabular}{c SS}
   \toprule
   $\xi$      & \mco{$\no{}$} & \mco{$\nv{}$} \\
   \midrule
   $10^{-1}$  &  1 &   4 \\
   $10^{-2}$  &  5 &   8 \\
   $10^{-3}$  & 16 &  56 \\
   $10^{-4}$  & 26 & 138 \\
   $10^{-5}$  & 29 & 208 \\
   $10^{-6}$  & 32 & 244 \\
   \midrule
   Full space & 39 & 263 \\
   \bottomrule
\end{tabular}
\end{table}
\begin{table}
\small
\centering
\caption{First 4 excited states of guanine with MLCC3 for descreasing CNTO thresholds.}
\label{t:Gua_4states}
\begin{tabular}{c cc cc cc cc}
   \toprule
   $\xi$ &
   \mc{2}{c}{State 1} &
   \mc{2}{c}{State 2} &
   \mc{2}{c}{State 3} &
   \mc{2}{c}{State 4}\\ \cmidrule(lr){2-3}\cmidrule(lr){4-5}\cmidrule(lr){6-7}\cmidrule(lr){8-9}
   &
   $\omega$ [eV] & $f\times 100$ &
   $\omega$ [eV] & $f\times 100$ &
   $\omega$ [eV] & $f\times 100$ &
   $\omega$ [eV] & $f\times 100$ \\
   \midrule
   CCSD      & 535.9067 & 3.20 & 538.4340 & 0.12 & 539.3858 & 0.05 & 539.6794 & 0.08 \\
   $10^{-2}$ & 534.8780 & 2.80 & 536.3546 & 0.08 & 537.7091 & 0.02 & 537.8040 & 0.00 \\
   $10^{-3}$ & 533.9879 & 2.43 & 535.1010 & 0.07 & 535.6097 & 0.11 & 536.1425 & 0.00 \\
   $10^{-4}$ & 533.5776 & 2.17 & 534.5033 & 0.06 & 534.7363 & 0.15 & 535.3402 & 0.01 \\
   $10^{-5}$ & 533.5184 & 2.12 & 534.3886 & 0.05 & 534.6080 & 0.15 & 535.1326 & 0.01 \\
   $10^{-6}$ & 533.5107 & 2.12 & 534.3704 & 0.05 & 534.5925 & 0.15 & 535.0691 & 0.02 \\
   \midrule
   CC3 & 533.5091 & 2.12 & 534.3599 & 0.05 & 534.5888 & 0.15 & 535.0139 & 0.02 \\
   \bottomrule
\end{tabular}
\end{table}
%
%
%
%
The results of Table \ref{t:Gua_4states} are plotted in Figure \ref{fig:cntoG}
in addition to the CCSD and CC3 results, depicted by the horizontal lines.
Increasing the active space improves the energies until the error is below
the expected error of the full CC3 method at a CNTO threshold of $10^{-4}$.
The oscillator strengths of the first and second state converge smoothly towards
their CC3 values,
however,
larger jumps are found for the third and fourth state.
These jumps are artifacts of the small active spaces,
the plots in the SI show a smooth convergence towards the CC3 values.
For the oscillator strengths the CCSD values have not been plotted
as horizontal lines because they would overload the plot,
and they coincide with the data points for $\xi=10^{-1}$.
\begin{table}
\centering
\caption{Timings in seconds to compute four core excited states from the oxygen atom of guanine
         at the CCSD, CC3 and MLCC3 level with decreasing CNTO threshold.
         Timings are given, averaged over the number of iterations when solving for
         $\bm{\tau}$, $\bm{\lambda}$, $\bm{R}$ and $\bm{L}$.
         Additionally, timings to construct the ground state density, $\bm{\DGS}$,
         left transition density, $\bm{\DL}$, and right transition density, $\bm{\DR}$,
         are reported.
         Note that the MLCC3 and CC3 timings only comprise the triples part.}
\label{t:Gua_time_4}
\begin{tabular}{l S SSSSS S}
   \toprule
   & \mc{1}{c}{CCSD} & \mc{5}{c}{MLCC3} & \mco{CC3}\\
   \cmidrule(lr){2-2} \cmidrule(lr){3-7} \cmidrule(lr){8-8}
   &  & \mco{$ 10^{-2}$} & \mco{$ 10^{-3}$}
      & \mco{$ 10^{-4}$} & \mco{$ 10^{-5}$} & \mco{$ 10^{-6}$} \\
   \midrule
   $\bm{\tau}$     & 22.8 & 0.08 & 2.75 & 133.79 &  704.15 & 1783.39 & 4180.2 \\
   $\bm{\lambda}$  & 44.2 & 0.13 & 5.68 & 270.98 & 1414.40 & 3408.42 & 8593.9 \\
   $\bm{R}$        & 38.7 & 0.10 & 1.23 &  30.89 &  149.33 &  342.51 &  700.2 \\
   $\bm{L}$        & 43.3 & 0.12 & 1.31 &  32.27 &  145.30 &  318.53 &  702.0 \\
   $\bm{\DGS}$     &  0.6 & 0.02 & 6.27 & 324.25 & 1658.26 & 4020.82 & 8765.3 \\
   $\bm{\DL}$      &  0.3 & 0.01 & 3.57 & 164.82 &  735.92 & 1662.66 & 3033.6 \\
   $\bm{\DR}$      &  1.2 & 0.06 & 7.39 & 330.58 & 1546.95 & 3647.72 & 7224.9 \\
   \bottomrule
\end{tabular}
\end{table}
Table \ref{t:Gua_time_4} shows the timings of one iteration of the most expensive
parts of the calculation of MLCC3 oscillator strengths.
For thresholds below $10^{-4}$ the CC3 contribution is negligible
when solving for ground and excited state amplitudes.
However,
the calculation of the EOM densities is already dominated by the CC3 part
at $\xi=10^{-3}$.
Compared to the timings for solving the amplitudes the densities are still
insignificant at a threshold of $10^{-3}$.
At $10^{-4}$ the CC3 contribution dominates all timings,
but compared to a full CC3 calculation the cost per iteration is reduced
by more than a factor of 30 for the ground state equations 20 for the excited states
(SI Table \ref{si-t:Gua_SU_xi}).
Even at $10^{-6}$ there is still a reduction of a factor of two,
despite most orbitals being included in the active space.

Comparing Table \ref{t:Gua_1state} and \ref{t:Gua_4states},
shows that the results with 20 occupied and 200 virtual orbitals are
slightly worse than the first excitation for $\xi=10^{-4}$,
although the latter includes only 6 more occupied but 62 less virtual orbitals.
Therefore,
we included calculations with a lower ratio between active virtual and occupied
orbitals.
Table \ref{t:Gua_1state_lower} shows the results for these calculations,
confirming that significantly less virtual orbitals are needed to obtain
almost identical results.
With 18 occupied and 130 virutal orbitals a speed up of up to 80 is achieved,
and with 20 occupied and 130 virtual orbitals the speed up is still around 50
(SI Table \ref{si-t:Gua_SU_o_to_v}).
\begin{table}
\centering
\caption{Calculations of a single core excited state of guanine from the oxygen atom
         at the CCSD, CC3 and MLCC3 level with varying sizes of the active space.
         Excitation energies, $\omega$, and oscillator strengths, $f$,
         as well as timings to construct the ground state density, $\bm{\DGS}$,
         left transition density, $\bm{\DL}$, and right transition density, $\bm{\DR}$,
         are reported.
         Additionally timings are given,
         averaged over the number of iterations when solving for
         $\bm{\tau}$, $\bm{\lambda}$, $\bm{R}$ and $\bm{L}$.
         Note that the MLCC3 and CC3 timings only comprise the triples
         part and that timings are given in seconds.}
\label{t:Gua_1state_lower}
\begin{tabular}{l SSSSSS}
   \toprule
   & \mc{6}{c}{MLCC3} \\
   \cmidrule(lr){1-1}\cmidrule(lr){2-7}
   $\no{}/\nv{}$ & \mco{16/160} & \mco{18/130} & \mco{18/150} & \mco{18/180} & \mco{20/130} & \mco{20/200} \\
   \midrule
   $\omega$ [eV] & 533.66 & 533.64 & 533.62 & 533.61 & 533.61 & 533.58 \\
   $f\times 100$ &   2.23 &   2.24 &   2.21 &   2.20 &   2.22 &   2.18 \\
   \midrule
   $\bm{\tau}$    & 40.09 & 28.09 & 41.22 & 70.31 & 38.32 & 125.74 \\
   $\bm{\lambda}$ & 70.96 & 53.86 & 76.78 & 130.22 & 73.14 & 238.02 \\
   $\bm{R}$       & 12.83 & 8.70 & 12.30 & 20.90 & 10.59 & 33.13 \\
   $\bm{L}$       & 13.43 & 9.69 & 13.23 & 22.59 & 11.88 & 36.58 \\
   $\bm{\DGS}$    & 83.67 & 69.90 & 103.69 & 175.06 & 94.52 & 312.30 \\
   $\bm{\DL}$     & 48.37 & 38.19 & 59.63 & 95.24 & 50.33 & 171.81 \\
   $\bm{\DR}$     & 90.03 & 75.51 & 111.44 & 182.68 & 95.11 & 320.64 \\
   \bottomrule
\end{tabular}
\end{table}

\begin{figure}
   \centering
   \includegraphics[width=\textwidth]{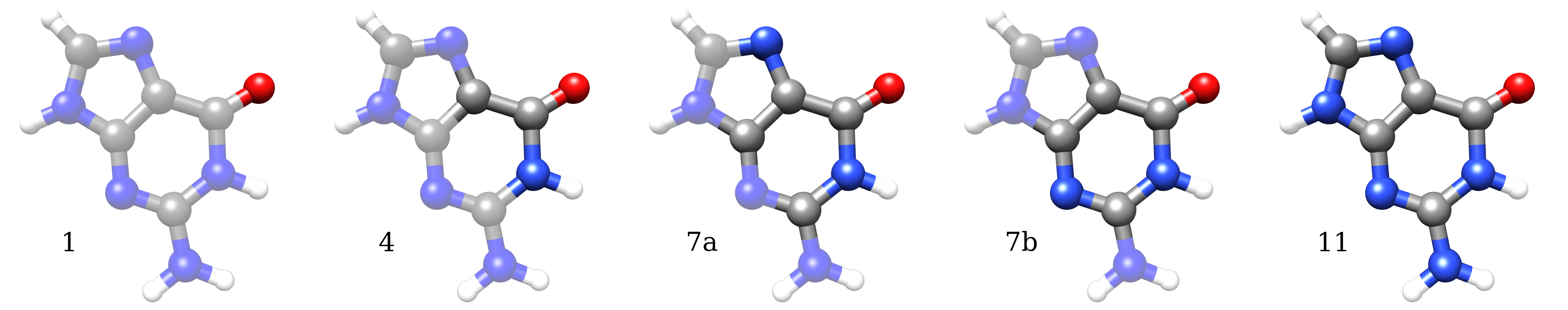}
   \caption{Geometry of guanine showing the active regions for which occupied
            Cholesky orbitals and PAOs have been constructed.
            The labels denote the number of active atoms and hydrogens are always inactive.}
   \label{fig:G_PAO_geom}
\end{figure}
We
have also performed
some calculations with Cholesky occupied orbitals and PAOs for the
virtual space.
The active atoms are shown in Figure \ref{fig:G_PAO_geom} as solid atoms
and $10^{-2}$ was used as threshold for the Choleksy decomposition of the AO density.
\begin{table}
\centering
\caption{Number of occupied and virtual orbitals in the active spaces
         constructed using Cholesky orbitals and PAOs for guanine.}
\label{t:pao_as}
\begin{tabular}{c SS}
   \toprule
   System label (Figure \ref{fig:G_PAO_geom})& \mco{$\no{}$} & \mco{$\nv{}$} \\
   \midrule
   1  &  5 &  26 \\
   4  & 23 &  92 \\
   7a & 33 & 158 \\
   7b & 33 & 158 \\
   11 & 39 & 245 \\
   \midrule
   Full space & 39 & 263 \\
   \bottomrule
\end{tabular}
\end{table}
\begin{table}
\small
\centering
\caption{First 4 excited states of Guanine with MLCC3 calculated with active spaces constructed from Cholesky orbitals and PAOs.}
\label{t:paoG_conv}
\begin{tabular}{r cc cc cc cc}
   \toprule
   &
   \mc{2}{c}{State 1} &
   \mc{2}{c}{State 2} &
   \mc{2}{c}{State 3} &
   \mc{2}{c}{State 4}\\ \cmidrule(lr){2-3}\cmidrule(lr){4-5}\cmidrule(lr){6-7}\cmidrule(lr){8-9}
   &
   $\omega$ [eV] & $f\times 100$ &
   $\omega$ [eV] & $f\times 100$ &
   $\omega$ [eV] & $f\times 100$ &
   $\omega$ [eV] & $f\times 100$ \\
   \midrule
   $1$  & 534.6250 & 2.81 & 537.0755 & 0.13 & 538.2042 & 0.39 & 538.2714 & 0.28 \\
   $4$  & 533.8427 & 2.37 & 534.9479 & 0.07 & 535.9413 & 0.05 & 536.0856 & 0.15 \\
   $7$a & 533.5901 & 2.17 & 534.6548 & 0.06 & 534.9312 & 0.16 & 535.4104 & 0.04 \\
   $7$b & 533.6082 & 2.19 & 534.6045 & 0.06 & 534.8809 & 0.16 & 535.4454 & 0.03 \\
   $11$ & 533.5110 & 2.12 & 534.4149 & 0.05 & 534.5932 & 0.15 & 535.0960 & 0.02 \\
   \midrule
   CC3 & 533.5091 & 2.12 & 534.3599 & 0.05 & 534.5888 & 0.15 & 535.01394 & 0.02 \\
   \bottomrule
\end{tabular}
\end{table}
The size of the active spaces are summarized in Table \ref{t:pao_as}
and the results are reported in Table \ref{t:paoG_conv}.
\begin{figure}[ht]
   \centering
   \begin{subfigure}{0.56\textwidth}
      \includegraphics[width=\textwidth]{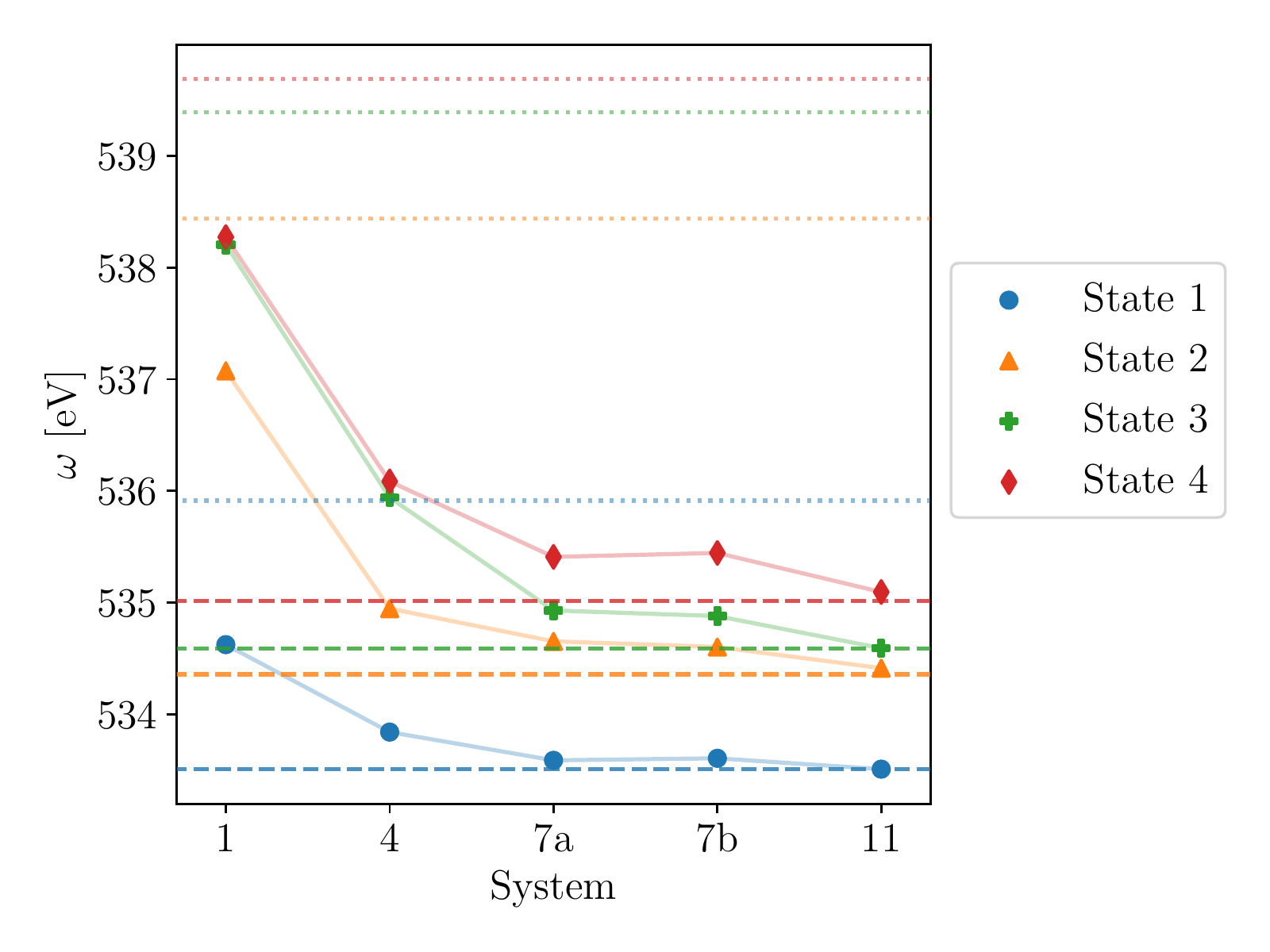}
   \end{subfigure}
   \begin{subfigure}{0.43\textwidth}
      \includegraphics[width=\textwidth]{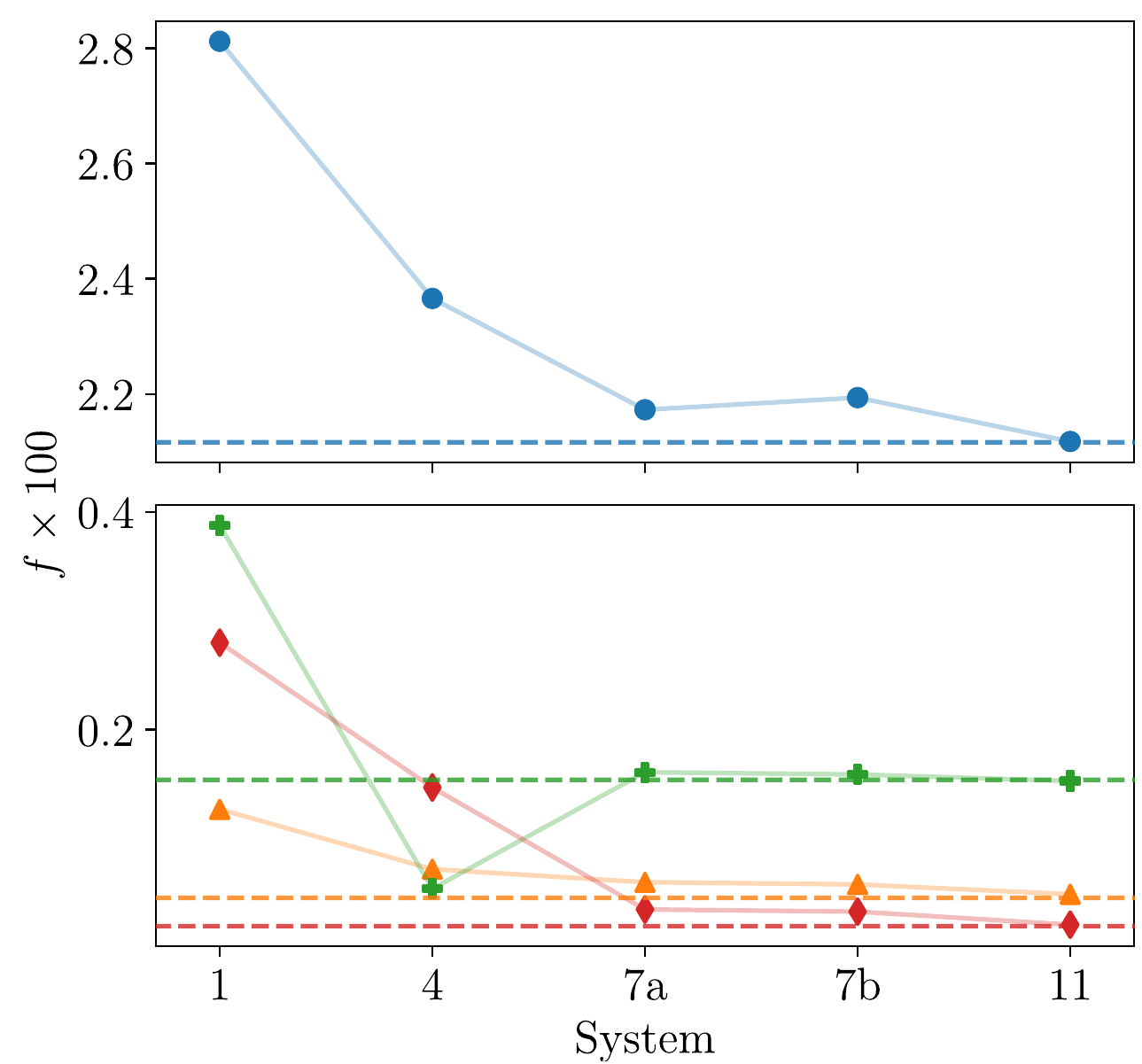}
   \end{subfigure}
   \caption{Convergence of the first four core excitation energies (left)
            and oscillator strengths (right) of guanine for the five
            active spaces in Figure \ref{fig:G_PAO_geom}.
            Dashed lines are the CC3 results and dotted lines denote the CCSD values.}
   \label{fig:paoG}
\end{figure}
As shown in Figure \ref{fig:paoG} the excitation energies are already significantly
improved when only the oxygen is included in the active space.
However, the size of the active spaces also increases much faster,
because all the atoms contribute to the $\pi$-system.
Despite the active spaces being larger,
the performance of the Cholesky/PAOs is worse than calculations with a similarly large
active space consisting of CNTOs.
The reason for the poor performance of these active spaces 
is that we split up the $\pi$-system.
Additionally, 
the CC3 excitation vectors consist of multiple similarly large amplitudes
which need to be described accurately by the active space.
An active space consisting of CNTOs is better suited to describe such excted states.

\subsection{Formaldehyde in water}

To investigate the scaling with the size of the inactive orbital space
we consider formaldehyde with several explicit water molecules.
The calculations were performed on two Intel Xeon-Gold 6138 using 40 threads.
Comparing excitation energy and oscillator strength
is not constructive for this system,
because CCSD and CC3 already almost coincide for the first excited state.
The geometry for formaldehyde with six water molecules
is reported in the SI;
it has been adapted from a geometry with 10 water molecules from Ref. \citenum{MLCC_CNTO}.
The other geometries are
generated by subsequently removing water molecules,
starting with the last one.
For a proper investigation of solvent effects,
randomized geometries would have to be extracted from a molecular dynamics
simulation and the results would have to be averaged.\cite{QMMM_ES}

For all calculations we used a aug-cc-pVTZ basis set
and the active space comprises 8 occupied and 136 virtual orbitals.
The sizes of the systems considered are summarized in Table \ref{t:size_fa}.
\begin{table}
\centering
\caption{Number of occupied and virtual orbitals for formaldehyde with
         increasing number of water molecules with aug-cc-pVTZ basis set.}
\label{t:size_fa}
\begin{tabular}{c cc}
   \toprule
   System & \mc{2}{c}{aug-cc-pVTZ} \\
   \cmidrule(lr){2-3}
    \#H$_2$O & \mco{$\NO{}$} & \mco{$\NV{}$} \\
   \midrule
    1 & 13 & 217 \\
    2 & 18 & 304 \\
    3 & 23 & 391 \\
    4 & 28 & 478 \\
    5 & 33 & 565 \\
    6 & 38 & 652 \\
   \bottomrule
\end{tabular}
\end{table}

\begin{figure}[ht]
   \centering
   \includegraphics[width=0.7\textwidth]{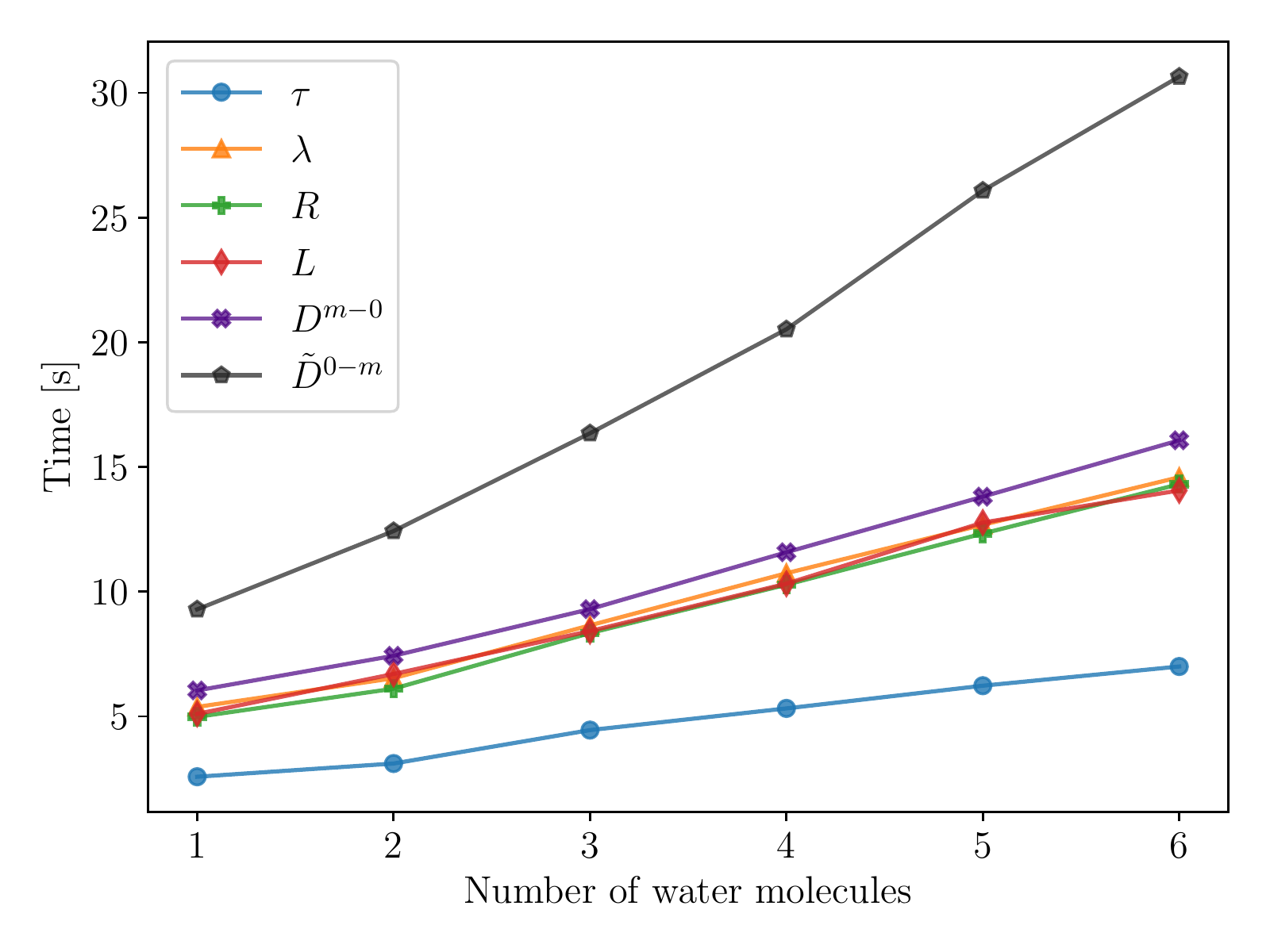}
   \caption{Average time to calculate one transition density or one
            iteration solving for $\bm{\tau}$, $\bm{\lambda}$, $\bm{L}$ and $\bm{R}$
            with increasing number of water molecules in the inactive space.}
   \label{fig:pao_water}
\end{figure}

Figure \ref{fig:pao_water} shows
the timing breakdown for the MLCC3 contribution in the calculation of EOM oscillator strengths.
As expected the timings for every quantity increase linearly with the number of
water molecules added to the system,
implying the terms scaling quadratically with the full system size
are negligible.

\subsection{Azobenzene}

In the aug-cc-pVDZ basis, azobenzene has 48 occupied and 364 virtual orbitals.
On two Intel Xeon E5-2699 v4 processors using 40 threads
a single iteration of the CC3 ground state equations takes 6 hours.
As the Jacobian transformations are twice as expensive per state,
a CC3 calculation of 10 excited states is costly.

\begin{figure}[h!]
   \centering
   \includegraphics[scale=0.5]{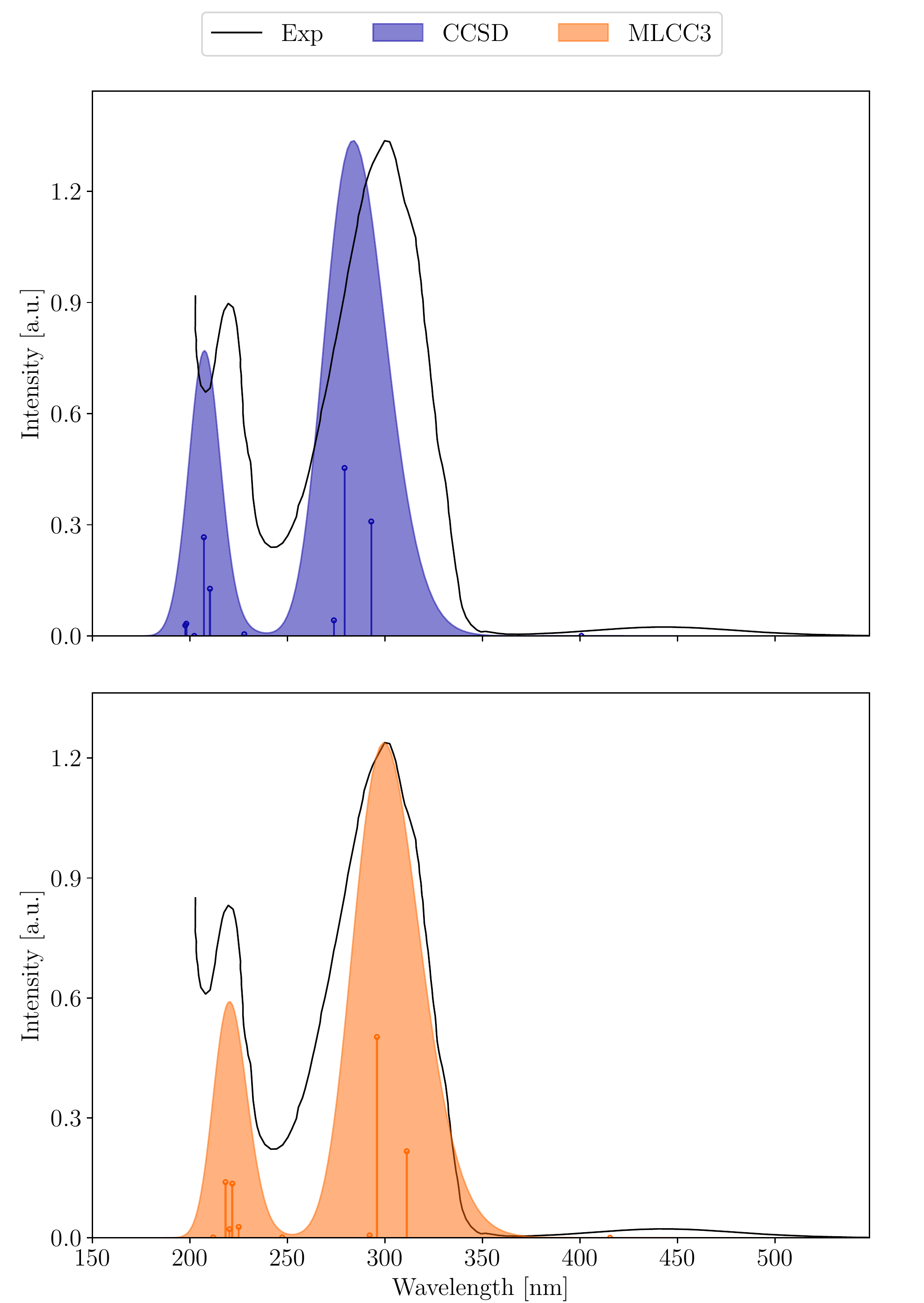}
   \caption{UV-VIS absorption spectrum of azobenezene calculated with CCSD
   and MLCC3 employing aug-cc-pVDZ basis set.\cite{aug_cc_pVXZ}
   The theoretical stick spectrum is broadened using Gaussian functions
   with fwhm of 0.5\,eV and the experimental data is
   taken from Ref. \citenum{azobenzene_exp}.}
   \label{fig:azobenzene}
\end{figure}

By using an active space containing 34 occupied and 238 virtual orbitals,
the time per iteration of the ground state equations reduced to 36 minutes.
In figure \ref{fig:azobenzene},
the spectra calculated at the CCSD and MLCC3 level of theory are shown
together with the experimental results.
While the CCSD results are significantly blue shifted,
the broadened MLCC3 values match very well with the experimental bands
at 300 nm and 220 nm.
The very broad band at around 450 nm is not reproduced,
but an almost dark excitation is found around 420 nm.
CCSD predicts this latter excitation to be at 400 nm instead.

\subsection{Betaine 30}

\begin{figure}[h!]
   \centering
   \includegraphics[width=0.4\textwidth]{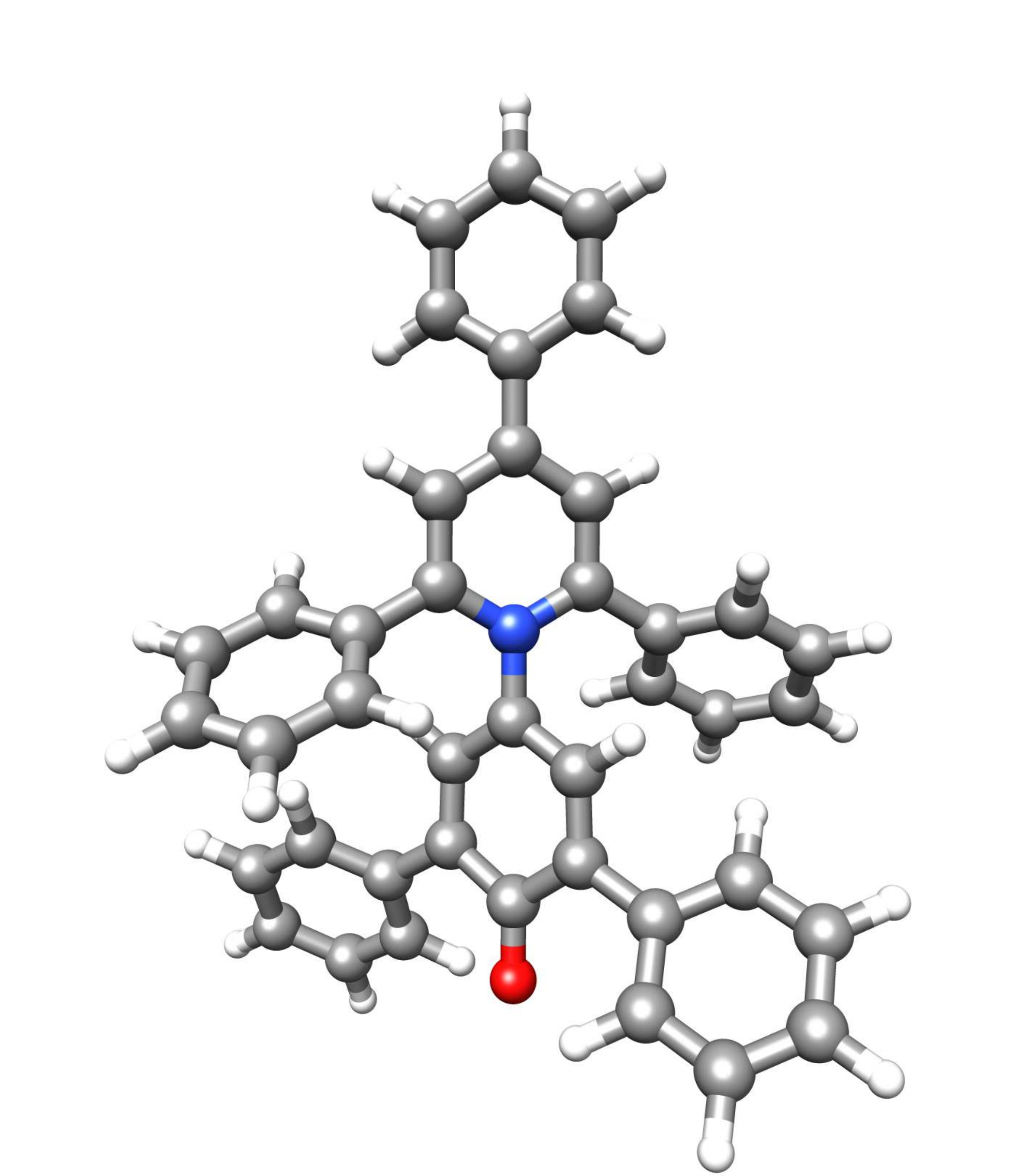}
   \caption{Geometry of the betaine 30.}
   \label{fig:betaine}
\end{figure}

To demonstrate the capabilities of our MLCC3 implementation, 
we consider the first core excitation from the oxygen atom in betaine 30.
The geometry is shown in Figure \ref{fig:betaine}.
The system comprises 145 occupied and 992 virtual orbitals
using a aug-cc-pCVDZ basis set for the oxygen atom,
aug-cc-pVDZ for carbon and nitrogen atoms and cc-pVDZ for hydrogen atoms.
\begin{table}[ht]
    \centering
    \begin{tabular}{c S[table-format=3.3] S[table-format=3.3] c c}
        \toprule
        & \mco{$\omega$ [eV]} & \mco{$f\times 100$} & $\no{}$ & $\nv{}$\\ \midrule
        CCSD  & 535.12 & 2.74 &  &  \\
        MLCC3 & 531.50 & 0.67 & 20 & 200 \\
        MLCC3 & 531.29 & 0.63 & 25 & 200 \\
        MLCC3 & 531.32 & 0.63 & 25 & 250 \\
        \bottomrule
    \end{tabular}
    \caption{First core excitation from the oxygen atom calculated at the CCSD level of theory 
             and MLCC3 with increasing number of CNTOs in the active space.}
    \label{tab:betaine}
\end{table}
In Table \ref{tab:betaine} we report the excitation energy and oscillator strengths for CCSD
and MLCC3 using three active CNTO spaces of increasing size.
Using CCSD both the excitation energy and especially the oscillator strength are
overestimated compared to the MLCC3 results.
Increasing the size of the active space from 20 occupied and 200 virtual CNTOs
to 25 occupied and 250 virtual orbitals, 
only changes the excitation energy by 0.3~eV.
Therefore,
we can assume that the MLCC3 results are within the expected error range of a full CC3 calculation.
\begin{table}[ht]
\centering
\caption{Timings in minutes to compute a core excited state
         from the oxygen atom of betaine 30
         at the CCSD and MLCC3 level with several active spaces.
         Timings are given, averaged over the number of iterations when solving for
         $\bm{\tau}$, $\bm{\lambda}$, $\bm{R}$ and $\bm{L}$.
         Additionally, timings to construct the ground state density, $\bm{\DGS}$,
         left transition density, $\bm{\DL}$, and right transition density, $\bm{\DR}$,
         are reported.
         Note that the MLCC3 and CC3 timings only comprise the triples part.}
\label{t:betaine_timings}
\begin{tabular}{l SS SSSSS}
   \toprule
   & \mco{CCSD} & \mc{3}{c}{MLCC3} \\
   \cmidrule(lr){2-2}\cmidrule(lr){3-5}
   $\no{}/\nv{}$ & & \mco{$20/200$} & \mco{$25/200$} & \mco{$25/250$} \\
   \midrule
   $\bm{\tau}$    &  73.2 & 3.1 &  5.6 & 10.7 \\
   $\bm{\lambda}$ & 143.5 & 6.4 & 12.0 & 21.4 \\
   $\bm{R}$       & 122.6 & 1.2 &  1.7 &  2.8 \\
   $\bm{L}$       & 130.8 & 1.3 &  1.9 &  3.2 \\
   $\bm{\DGS}$    &   0.5 & 8.0 & 14.6 & 28.6 \\
   $\bm{\DL}$     &   0.5 & 5.2 &  9.4 & 18.3 \\
   $\bm{\DR}$     &   1.0 & 9.1 & 16.5 & 31.2 \\
   \bottomrule
\end{tabular}
\end{table}

Due to the significant size of the system the time spent calculating 
the contribution of the triple excitations is small compared to the timings of CCSD,
as shown in Table \ref{t:betaine_timings}.
For the densities the triples contribution dominates, 
however,
the time used to construct densities is still small compared to
determining the ground and excited states.

\section{Conclusion}\label{Conclusion}

The multilevel CC3 method provides a framework,
with which intensive molecular properties can be calculated
at an accuracy approaching that of the CC3 method.
For sufficiently large inactive spaces the computational cost
will tend towards that of CCSD.
Compared to Cholesky PAOs,
CNTOs provide smaller orbital spaces without sacrificing accuracy.
However,
the cost of constructing CNTOs is significant,
as the CCSD ground and excited state equations need to be solved.

There is some ambiguity regarding the selection of the active space using CNTOs.
We can either specify the number of occupied and virtual orbitals explicitly
or use a cutoff, $\xi$,
and include the orbitals whose eigenvalues sum up to $1-\xi$.
The first approach gives great flexibility,
but several calculations are typically needed to confirm that the excitation
energies actually converged.
Using a cutoff on the other hand is a more blackbox approach,
as $\xi=10^{-4}$ gives accurate results,
but the active spaces can become larger than required.
Further benchmarking, especially on larger systems,
is needed to obtain a rule of thumb for the selection of an active space.


Two bottlenecks were identified that limit the size of the systems we can treat:
First, the convergence behaviour of the solvers is diminished,
due to the change of the orbital basis.
The start guess could be improved by transforming the CCSD amplitudes
from canonical MOs to the CNTO basis.

For large systems with several hundered to a thousand MOs,
CCSD becomes a bottleneck and
another layer could be introduced at the CCS level of theory.
For the multilevel CC3 model with CC3 in CCSD in CCS,
it has to be investigated how the orbital space is set up effectively,
as NTOs obtained from CCS will not provide a suitable active space.
One possibility could be the approximated CNTOs introduced by Baudin and Kristensen,
or CNTOs obtained from a MLCCSD calculation.\cite{cornflex}


\begin{acknowledgement}

We acknowledge computing resources through UNINETT Sigma2
- the National Infrastructure for High Performance Computing
and Data Storage in Norway, through project number NN2962k.
We acknowledge funding from the Marie Sk{\l}odowska-Curie European Training Network
``COSINE -- COmputational Spectroscopy In Natural sciences and Engineering'',
Grant Agreement No. 765739 and
the Research Council of Norway through FRINATEK projects 263110, CCGPU,
and 275506, TheoLight.

\end{acknowledgement}


\clearpage


\bibliography{mlcc3}

\begin{mcitethebibliography}{87}
\bibitem[Lin and Truhlar(2007)Lin, and Truhlar]{QMMM_review_1}
Lin,~H.; Truhlar,~D.~G. {QM/MM: what have we learned, where are we, and where
  do we go from here?} \emph{Theor Chem Acc} \textbf{2007}, \emph{117},
  185--199\relax
\mciteBstWouldAddEndPuncttrue
\mciteSetBstMidEndSepPunct{\mcitedefaultmidpunct}
{\mcitedefaultendpunct}{\mcitedefaultseppunct}\relax
\EndOfBibitem
\end{mcitethebibliography}

\end{document}



\begin{suppinfo}

Here we report equations to construct the MLCC3 ground state residual,
Jacobian transformations and transition densities.
For a concise notation we define
\begin{align}
   \Del{ai,bj} &= 1 + \delta_{ij}\delta_{ab} \\
   \nR{ab}{ij} &= \Del{ai,bj} R^{ab}_{ij}.
\end{align}
For a given covariant doubles amplitude or residual $X^{ab}_{ij}$ the
contravariant quantity is defined as
\begin{equation}
   \tilde{X}^{ab}_{ij} = 2X^{ab}_{ij} - X^{ba}_{ij}
\end{equation}
and for a triples amplitude $X^{abc}_{ijk}$
\begin{equation}
   \tilde{X}^{abc}_{ijk} =
   4X^{abc}_{ijk} - 2X^{bac}_{jik} - 2X^{cba}_{kji}
   - 2X^{acb}_{ikj} + X^{cab}_{kij} + X^{bca}_{jki}
\end{equation}

The contributions to the contravariant ground state residual $\bm{\tOm{}{}}$ are listed below.
\begin{align}
   &\ta{abc}{ijk} = -(\eps^{abc}_{ijk})^{-1} P^{abc}_{ijk}
                     \Big(\sum_D \ta{aD}{ij} g_{bDck} -
                     \sum_l \ta{ab}{iL} g_{Ljck}\Big)
   \\
   &\tOm{a}{i} \peq \ssum{bc}{jk} \tta{abc}{ijk} g_{jbkc}
   \\
   &\tOm{ab}{ij} \peq{} P^{ab}_{ij} \ssum{c}{k} \tta{abc}{ijk} F_{kc}
   \\
   &\tOm{ab}{iL} \meq{} P^{ab}_{iL} \ssum{c}{jk} \tta{abc}{ijk} g_{jLkc}
   \\
   &\tOm{aD}{ij} \peq{} P^{aD}_{ij} \ssum{bc}{jk} \tta{abc}{ijk} g_{Dbkc}
\end{align}

The Jacobian transformation of a trial vector $\bm{R}$ consists of the following terms,
where $\bm{\tilde{\rho}}$ denotes the contravariant of the transformed vector.
\begin{align}
   &\Ups{kc} = \ssum{D}{L} (2g_{kcLD}-g_{kDLc}) R^D_L \\
   &\Ups{bDck} = \sum_E R^E_k g_{bDcE}
               - \sum_M \big(R^b_M g_{MDck} + R^c_M g_{bDMk}\big) \\
   &\Ups{Ljck} = \sum_E \big(R^E_j g_{LEck} + R^E_k g_{LjcE}\big)
               - \sum_M R^c_M g_{LjMk}
   \\
   &\Ra{abc}{ijk} = -\frac{1}{\eps^{abc}_{ijk} - \omega} P^{abc}_{ijk}
      \Big(  \sum_D \nR{aD}{ij} g_{bDck}   - \sum_L \nR{ab}{iL} g_{Ljck}
           + \sum_D \ta{aD}{ij} \Ups{bDck} - \sum_L \ta{ab}{iL} \Ups{Ljck} \Big)
   \\
   &\trho{a}{i} \peq \ssum{bc}{jk} \tR{abc}{ijk} g_{jbkc}
   \\
   &\trho{ab}{ij} \peq{} \Del{aibj}^{-1} P^{ab}_{ij} \ssum{c}{k}
         \Big(\tR{abc}{ijk} F_{kc} + \tta{abc}{ijk} \Ups{kc}\Big)
   \\
   &\trho{ab}{iL} \meq{} \Del{aibL}^{-1}P^{ab}_{iL}\Big(
         \ssum{c}{jk} \tR{abc}{ijk} g_{jLkc}
      +  \ssum{cD}{jk} \ta{abc}{ijk} g_{jDkc} R^D_L\Big)
   \\
   &\trho{aD}{ij} \peq{} \Del{aiDj}^{-1}P^{aD}_{ij}\Big(
        \ssum{bc}{k} \tR{abc}{ijk} g_{Dbkc}
      - \ssum{bc}{kL} \tta{abc}{ijk} g_{Lbkc} R^D_L\Big)
\end{align}

The transformation of a trial vector $\bm{L}$ with the transpose of the Jacobian
is calculated as follows,
where $\bm{\sigma}$ denotes the contravariant of the transformed vector.
\begin{align}
   &L^{abc}_{ijk} = \frac{1}{\omega - \eps^{abc}_{ijk}} \, P^{abc}_{ijk}
                  \Big(L^a_i g_{jbkc} + L^{ab}_{ij} F_{kc}
                  - \sum_L L^{ab}_{Lk} g_{iLjc} + \sum_D L^{aD}_{jk} g_{ibDc}\Big)
   \\
   &\sigma^D_L \peq
      \ssum{abc}{ijk} \tta{abc}{ijk} L^{ab}_{ij} (2g_{kcLD}-2g_{kDLc})
   +  \ssum{abc}{ijk} \tta{abc}{ijk} g_{Lbkc} L^{aD}_{ij}
   +  \ssum{abc}{ijk} \tta{abc}{ijk} g_{jDkc} L^{ab}_{iL}
   \\
   &\sigma^D_l \peq
     \ssum{abcE}{ij} \tL{abc}{ijl} t^{aE}_{ij} g_{bEcD}
   - \ssum{abc}{ijM} \tL{abc}{ijl} t^{ab}_{iM} g_{MjcD}
   - \ssum{abc}{ikM} \tL{abc}{ilk} t^{ab}_{iM} g_{MDck}
   \\
   &\sigma^d_L \peq
     \ssum{ab}{ijkM} \tL{abd}{ijk} t^{ab}_{iM} g_{MjLk}
   - \ssum{abE}{ijk} \tL{abd}{ijk} t^{aE}_{ij} g_{LkbE}
   - \ssum{acE}{ijk} \tL{adc}{ijk} t^{aE}_{ij} g_{LEck}
   \\
   &\sigma^{aD}_{ij} \peq P^{aD}_{ij} \ssum{bc}{k} \tL{abc}{ijk} g_{bDck}\\
   &\sigma^{ab}_{iL} \meq P^{ab}_{iL} \ssum{c}{kl} \tL{abc}{ijk} g_{Ljck}
\end{align}
Using Cholesky decomposition the integral $g_{PQRS}$ are decomposed into
$\sum_\chi {L^\chi_{PQ} L^\chi_{RS}}$ reducing the memory requirements for
the integrals and intermediates.

The following equations contain the CC3 contribution to the left transition density $\DL$
\begin{align}
   &\DL_{kl} \meq \ssum{abc}{ij} \half \tL{abc}{ijl} \ta{abc}{ijk}
   \\
   &\DL_{ld} \peq \ssum{ab}{ij} L^{ab}_{ij} \tta{abd}{ijl} \\
   &\DL_{LD} \meq \ssum{abc}{ijk} \tL{abc}{ijk} \ta{ac}{iL} \ta{bD}{jk}
   \\
   &\DL_{cd} \peq \ssum{ab}{ijk} \half \tL{abc}{ijk} \ta{abd}{ijk}
\end{align}
The ground state density $D^{0-0}$ is obtained if $\tL{}{}$ is substituted by $\tla{}{}$.

Finally the CC3 terms for the right transition density, $\DR$.
\begin{equation}
\begin{split}
    &\DR_{Kl} \meq \ssum{abc}{ij} \tla{abc}{ijl} \Ra{a}{i} \ta{bc}{jK} \\
    &\DR_{kl} \meq  \half\ssum{abc}{ij} \tla{abc}{ijl} \Ra{abc}{ijk}
    \\
    &\DR_{ld} \peq \ssum{abc}{ijk} \tla{abc}{ijk} \Ra{a}{i}
    \big(\ta{bcd}{jkl} - \ta{bcd}{jlk}\big)
    + \ssum{ab}{ij} \tla{ab}{ij} \tR{abd}{ijl}
    \\
    &\DR_{LD} \peq  \half \ssum{abc}{ijk} \tla{abc}{ijk} \Ra{ab}{ij} (2 \ta{cD}{kL}-\ta{cD}{Lk})
    -\ssum{abc}{ijk} \tla{abc}{ijk} \big(\Ra{ac}{iL} \ta{bD}{jk} + \Ra{aD}{ik} \ta{bc}{jL}\big)
    \\
    &\DR_{lD} \meq \half \ssum{abc}{ijk} \tla{abc}{ijk}  \ta{abc}{ijl} \Ra{D}{k}
    \\
    &\DR_{Ld} \meq \half \ssum{abc}{ijk} \tla{abc}{ijk} \ta{abd}{ijk} \Ra{c}{L}
    \\
    &\DR_{ck} \peq \half \ssum{ab}{ij} \tla{abc}{ijk} \Ra{ab}{ij}
    \\
    &\DR_{cD} \peq \ssum{ab}{ijk} \tla{abc}{ijk} \Ra{a}{i} \ta{bD}{jk} \\
    &\DR_{cd} \peq \ssum{ab}{ijk} \half \tla{abc}{ijk} \Ra{abd}{ijk}
    \\
    &\DR_{LL} \peq \frac{1}{6} \ssum{abc}{ijk} 2\tla{abc}{ijk} \Ra{abc}{ijk}
    \\
    &\DR_{pq} \meq \frac{1}{6} \ssum{abc}{ijk} \tla{abc}{ijk} \Ra{abc}{ijk} D^{0-0}_{pq}
\end{split}
\end{equation}

\clearpage

\subsection*{Guanine}

Table \ref{t:Gua_4_systematic} summarizes the results from calculations
using MLCC3 with active spaces where $\nv{} = 10\no{}$.
These results are also visualized in Figure \ref{fig:Gua_4_systematic}
showing a smooth convergence of both excitation energies and oscillator strengths
towards the CC3 results.

\begin{table}[h!]
   \centering
   \caption{Excitation energies and oscillator strengths for the first four
            excited states of Guanine with MLCC3 for active spaces with increasing size.
            For the MLCC3 values the number of occupied and virtual orbitals
            is reported in the left column.
            The total system contains 39 occupied and 263 virtual orbitals.}
   \label{t:Gua_4_systematic}
   \begin{tabular}{c cc cc cc cc}
      \toprule
      System &
      \mc{2}{c}{State 1} & \mc{2}{c}{State 2} & \mc{2}{c}{State 3} & \mc{2}{c}{State 4} \\
      \cmidrule(lr){1-1} \cmidrule(lr){2-3} \cmidrule(lr){4-5}  \cmidrule(lr){6-7} \cmidrule(lr){8-9}
      CCSD & 535.91 & 3.26 & 538.44 & 0.12 & 539.39 & 0.05 & 539.68 & 0.07 \\
      \phantom{1}8/\phantom{1}80  & 534.13 & 2.45 & 535.20 & 0.06 & 536.22 & 0.10 & 536.34 & 0.00 \\
      10/100 & 533.94 & 2.39 & 534.99 & 0.06 & 535.79 & 0.11 & 536.08 & 0.00 \\
      13/130 & 533.79 & 2.30 & 534.78 & 0.06 & 535.43 & 0.13 & 535.77 & 0.01 \\
      15/150 & 533.73 & 2.26 & 534.66 & 0.06 & 535.18 & 0.14 & 535.64 & 0.01 \\
      18/180 & 533.66 & 2.21 & 534.55 & 0.05 & 534.94 & 0.14 & 535.44 & 0.01 \\
      20/200 & 533.62 & 2.18 & 534.49 & 0.05 & 534.83 & 0.16 & 535.32 & 0.01 \\
      24/240 & 533.55 & 2.14 & 534.41 & 0.05 & 534.69 & 0.17 & 535.14 & 0.02 \\
      CC3 & 533.51 & 2.12 & 534.36 & 0.05 & 534.59 & 0.15 & 535.01 & 0.02 \\
      \bottomrule
   \end{tabular}
\end{table}

\begin{figure}[ht]
   \centering
   \begin{subfigure}{0.555\textwidth}
      \includegraphics[width=\textwidth]{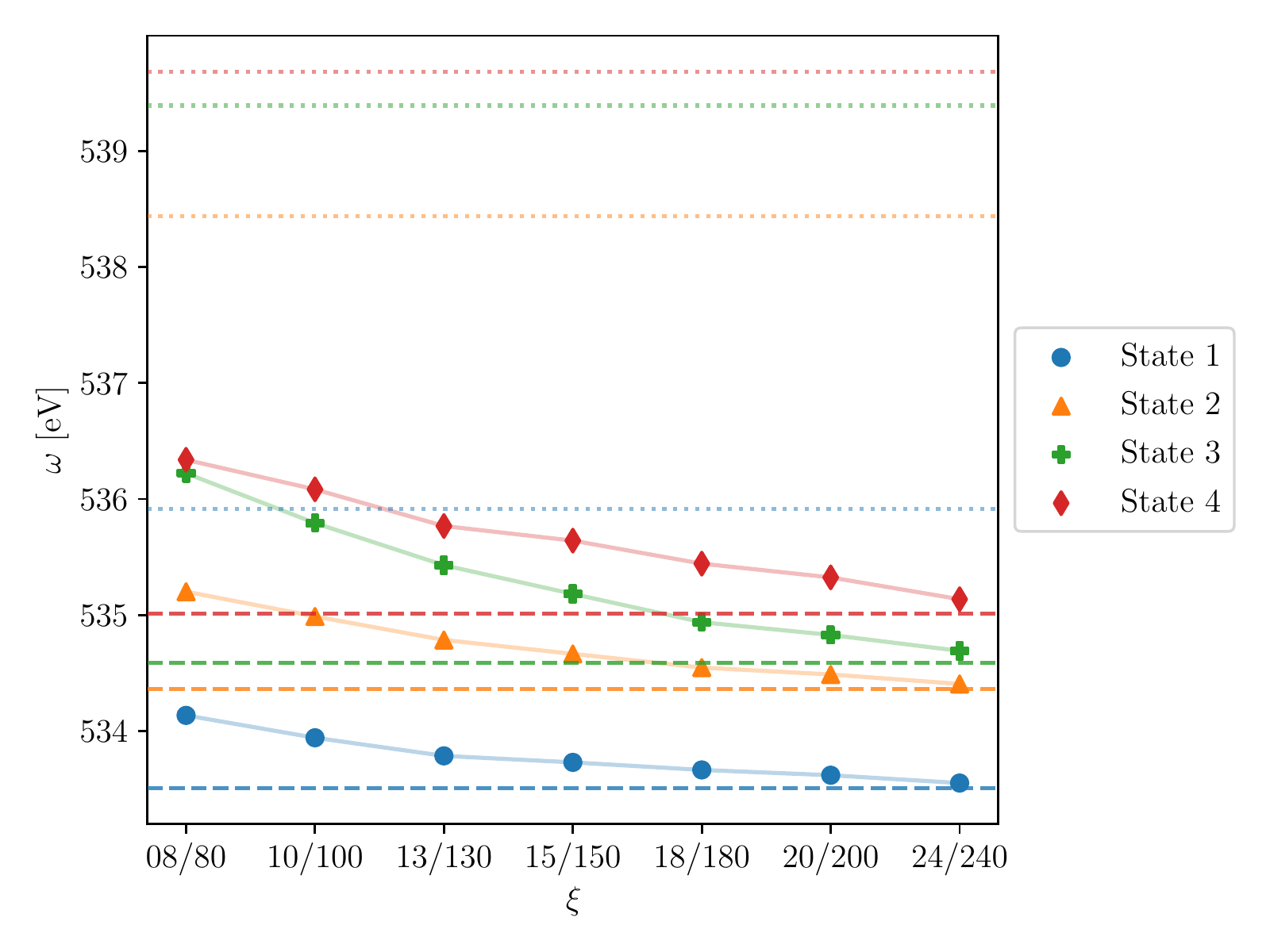}
   \end{subfigure}
   \begin{subfigure}{0.435\textwidth}
      \includegraphics[width=\textwidth]{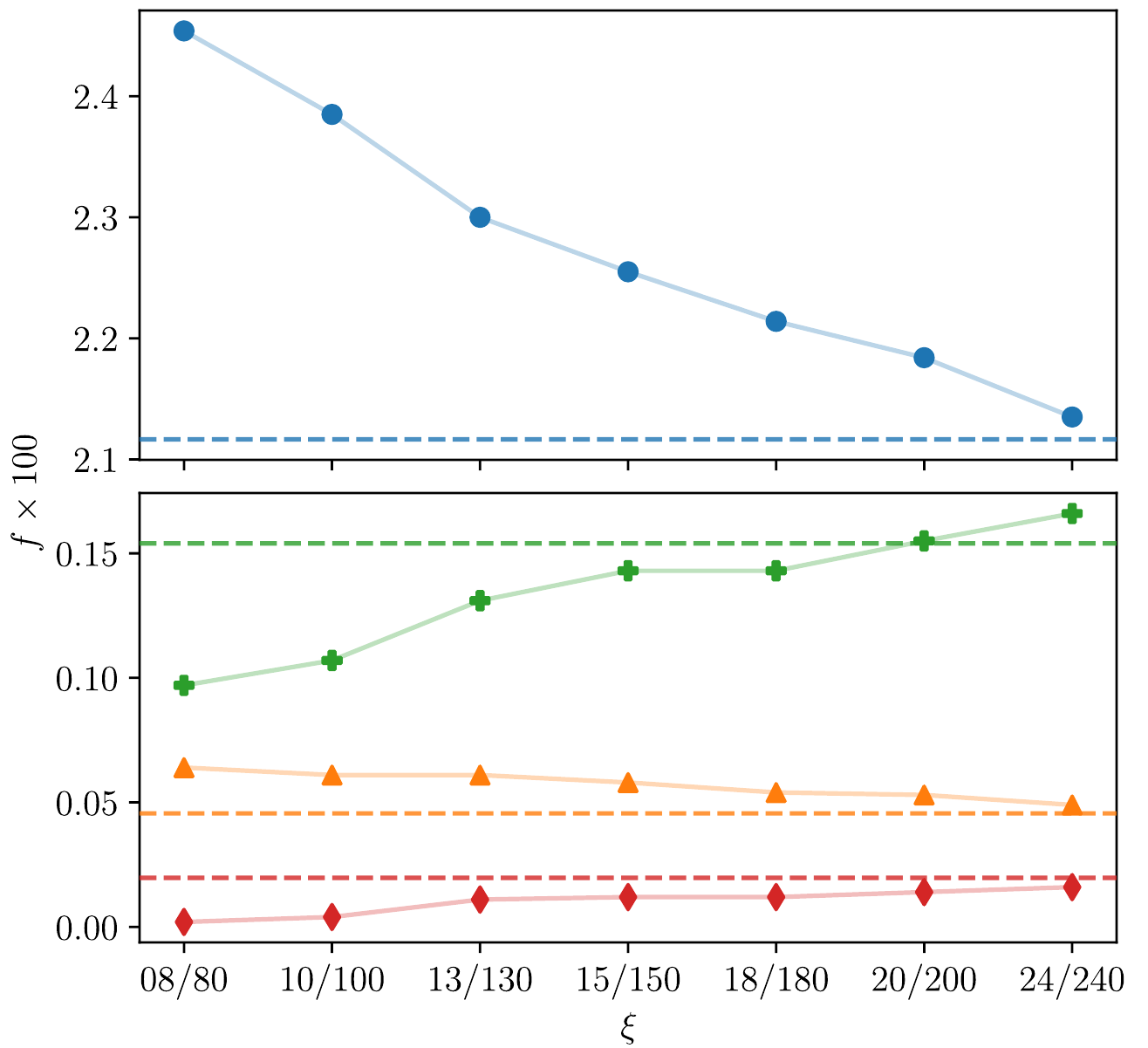}
   \end{subfigure}
   \caption{Convergence of the first four core excitation energies (left)
            and oscillator strengths (right) of guanine with the size
            of the active space.
            Dashed lines are the CC3 results and dotted lines denote the CCSD values.}
   \label{fig:Gua_4_systematic}
\end{figure}

\begin{table}
\centering
\caption{Speed up of MLCC3 compared to canonical CC3
         for the calculation of four core excited states of guanine.
         Speed ups calculated according to equations
         47 and 48 in the main document.
         The first part shows the speed up for terms that scale asymptotically
         as $\mcl{O}(\NV{}\nv{3}\no{3})$ while the second part summarizes the
         speed up for terms with a cost of $\mcl{O}(\NV{}\nv{3}\no{2})$.}
\label{t:Gua_SU_xi}
\begin{tabular}{l c
                S[table-number-alignment = center]
                S[table-number-alignment = center]
                S[table-number-alignment = center]
                S[table-number-alignment = center]}
   \toprule
   $\xi$ && \mco{$10^{-3}$} & \mco{$10^{-4}$} & \mco{$10^{-5}$} & \mco{$10^{-6}$} \\
   \midrule
   $\bm{\tau}$     && 1520.1 & 31.2 & 5.9 & 2.3 \\
   $\bm{\lambda}$  && 1505.2 & 31.6 & 6.0 & 2.5 \\
   $\bm{\DGS}$     && 1397.9 & 27.0 & 5.3 & 2.2 \\
   $\bm{\DL}$      &&  849.7 & 18.4 & 4.1 & 1.8 \\
   $\bm{\DR}$      &&  977.8 & 21.9 & 4.7 & 2.0 \\
   \cmidrule{1-1}
   $S^{GS}_{theo}$ && 1500.2 & 23.4 & 4.9 & 2.3 \\
   \midrule
   $\bm{R}$        &&  537.8 & 21.4 & 4.4 & 1.9 \\
   $\bm{L}$        &&  502.8 & 20.4 & 4.5 & 2.1 \\
   \cmidrule{1-1}
   $S^{ES}_{theo}$ &&  615.4 & 15.6 & 3.7 & 1.9 \\
   \bottomrule
\end{tabular}
\end{table}

\begin{table}
\centering
\caption{Speed up of MLCC3 compared to canonical CC3
         for the calculation of a single core excited states of guanine.
         Speed ups calculated according to equations
         47 and 48 in the main document.
         The first part shows the speed up for terms that scale asymptotically
         as $\mcl{O}(\NV{}\nv{3}\no{3})$ while the second part summarizes the
         speed up for terms with a cost of $\mcl{O}(\NV{}\nv{3}\no{2})$.}
\label{t:Gua_SU_o_to_v}
\begin{tabular}{l c
                S[table-number-alignment = center]
                S[table-number-alignment = center]
                S[table-number-alignment = center]
                S[table-number-alignment = center]
                S[table-number-alignment = center]
                S[table-number-alignment = center]}
   \toprule
   $\no{}$ && \mco{16} & \mco{18} & \mco{18} & \mco{18} & \mco{20} & \mco{20} \\
   $\nv{}$ && \mco{160} & \mco{130} & \mco{150} & \mco{180} & \mco{130} & \mco{200} \\
   \midrule
   $\bm{\tau}$     && 55.4 & 79.1 & 53.9 & 31.6 & 57.9 & 17.7 \\
   $\bm{\lambda}$  && 58.6 & 77.2 & 54.1 & 31.9 & 56.8 & 17.5 \\
   $\bm{\DGS}$     && 61.5 & 73.6 & 49.6 & 29.4 & 54.5 & 16.5 \\
   $\bm{\DL}$      && 51.5 & 61.4 & 41.6 & 25.4 & 48.8 & 14.5 \\
   $\bm{\DR}$      && 48.4 & 61.3 & 39.2 & 24.6 & 46.5 & 13.6 \\
   \cmidrule{1-1}
   $S^{GS}_{theo}$ && 64.3 & 84.2 & 54.8 & 31.7 & 61.4 & 16.9 \\
   \midrule
   $\bm{R}$        && 23.5 & 34.7 & 24.6 & 14.4 & 28.5 & 9.1 \\
   $\bm{L}$        && 23.7 & 32.8 & 24.0 & 14.1 & 26.8 & 8.7 \\
   \cmidrule{1-1}
   $S^{ES}_{theo}$ && 26.4 & 38.9 & 25.3 & 14.6 & 31.5 & 8.6 \\
   \bottomrule
\end{tabular}
\end{table}

\clearpage

\subsection*{Geometries}

Here we list the geometries of the molecules used in the calculations
presented in the application section of the paper.

\begin{table}
   \centering
   \caption{Geometry of guanine in {\AA}ngstr{\o}m.}
   \begin{tabular}{c S[table-format=3.6] S[table-format=3.6] S[table-format=3.6]}
      \toprule
      Atom & \mc{1}{c}{x} & \mc{1}{c}{y} & \mc{1}{c}{z} \\
      \midrule
      O  &   2.400416 &   1.186125 &  0.000000 \\
      N  &  -2.164547 &   0.729374 &  0.000000 \\
      C  &  -1.823669 &   2.066210 &  0.000000 \\
      N  &  -0.540931 &   2.253857 &  0.000000 \\
      C  &   0.000000 &   0.987621 &  0.000000 \\
      C  &   1.368570 &   0.557966 &  0.000000 \\
      N  &   1.424157 &  -0.869681 &  0.000000 \\
      C  &   0.356682 &  -1.723002 &  0.000000 \\
      N  &   0.627528 &  -3.053172 &  0.000000 \\
      N  &  -0.882076 &  -1.320799 &  0.000000 \\
      C  &  -0.996160 &   0.028079 &  0.000000 \\
      H  &  -2.573196 &   2.842649 &  0.000000 \\
      H  &   2.367552 &  -1.232740 &  0.000000 \\
      H  &   1.561780 &  -3.418613 &  0.000000 \\
      H  &  -0.152003 &  -3.687334 &  0.000000 \\
      H  &  -3.088910 &   0.328741 &  0.000000 \\
      \bottomrule
   \end{tabular}
\end{table}

\begin{table}
   \footnotesize
   \centering
   \caption{Geometry of formaldehyde with 6 explicit water molecules in {\AA}ngstr{\o}m.
   Adapted from a geometry with 10 water molecules from Ref. \citenum{MLCC_CNTO}.}
   \begin{tabular}{c S[table-format=3.7] S[table-format=3.7] S[table-format=3.7]}
      \toprule
      Atom & \mc{1}{c}{x} & \mc{1}{c}{y} & \mc{1}{c}{z} \\
      \midrule
      C &    0.24155 &  -0.26233 &   0.50653 \\
      O &    1.08878 &  -0.29115 &   1.39233 \\
      H &   -0.09479 &  -1.18677 &   0.01481 \\
      H &   -0.21855 &   0.68784 &   0.17709 \\
      O &    1.67506 &   2.52513 &   1.36591 \\
      H &    1.64135 &   1.57595 &   1.59689 \\
      H &    1.72610 &   2.51833 &   0.38456 \\
      O &    1.75008 &  -0.31632 &  -1.77458 \\
      H &    1.57721 &   0.65833 &  -1.80806 \\
      H &    2.61183 &  -0.36928 &  -1.33501 \\
      O &    1.48197 &  -3.01167 &   1.26413 \\
      H &    2.11276 &  -3.43941 &   1.86016 \\
      H &    1.49199 &  -2.05596 &   1.51182 \\
      O &   -2.40632 &  -1.19477 &   0.86990 \\
      H &   -2.04899 &  -2.10772 &   0.67851 \\
      H &   -3.05560 &  -1.30821 &   1.57778 \\
      O &   -0.89752 &   3.33556 &   1.62275 \\
      H &    0.06715 &   3.09459 &   1.64454 \\
      H &   -0.91078 &   4.27966 &   1.83820 \\
      O &   -1.16871 &   2.71932 &  -1.00701 \\
      H &   -1.22105 &   2.98736 &  -0.05388 \\
      H &   -1.77348 &   1.94185 &  -1.10382 \\
      \bottomrule
   \end{tabular}
\end{table}

\end{suppinfo}

\clearpage


\bibliography{mlcc3}